\documentclass[a4paper,11pt]{article}
\pdfoutput=1
\usepackage{amssymb}
\usepackage{epsfig}
\usepackage{graphicx}

\makeatletter

\@addtoreset{equation}{section}
\makeatother
\def\be{\begin{equation}}
\def\ee{\end{equation}}
\def\bea{\begin{eqnarray}}
\def\eea{\end{eqnarray}}

\def\({\left(}
\def\){\right)}
\def\<{\left<}
\def\>{\right>}

\def\tr{{\mbox{tr}}}
\def\be{\begin{equation}}
\def\ee{\end{equation}}
\def\bea{\begin{eqnarray*}}
\def\eea{\end{eqnarray*}}
\def\ben{\begin{eqnarray}}
\def\een{\end{eqnarray}}
\def\({\left(}
\def\){\right)}
\def\<{\left<}
\def\>{\right>}
\def\!{\right|}
\def\|{\left|}

\def\[{\left[}
\def\]{\right]}

\def\+{\bar}
\def\mb{\mathbb}
\def\tr{{\mbox{tr}}}

\def\A{{\cal{A}}}
\def\B{{\cal{B}}}

\def\Ordo{{\cal{O}}}
\def\Ordo{{\cal{O}}}

\def\R{{\mb{R}}}

\def\ell{\ell}

\def\h{\widehat}

\begin{document}

\begin{titlepage}
\vskip1cm
\begin{flushright}
UOSTP {\tt 140301}
\end{flushright}
\vskip 2.25cm
\centerline{\Large
\bf Elliptic genera of monopole strings
}
\vskip 1.25cm \centerline{  Dongsu Bak and  Andreas Gustavsson}
\vspace{1cm} \centerline{  Physics Department,
University of Seoul, { Seoul 130-743, Korea}} 
\vskip 0.75cm \centerline{(\tt dsbak@uos.ac.kr, agbrev@gmail.com)} 
\vspace{2.5cm}
\centerline{\bf Abstract} \vspace{0.75cm} \noindent
We obtain elliptic genera of monopole string in 5d MSYM. We find agreement with
the corresponding TST-dual dyonic-instanton single particle indices in 1110.2175. We make use of (2,2) 
superconformal algebra and its spectral flow, and the agreement can therefore be taken as 
evidence that monopole string (4,4) sigma models are exactly quantum superconformal.

\vspace{1.75cm}

\end{titlepage}

\section{Introduction}

In the recent development of M5 brane theory, the proposal of the M5/D4 correspondence \cite{Douglas:2010iu, Lambert:2010iw} plays 
a central role despite some critical difficulties. The proposal says that a system of M5 branes compactified on the M-theory circle is dual to the dimensionally reduced system of D4 branes whose worldvolume dynamics is described  by 5d maximally supersymmetric YM (MSYM) theory. The KK modes after the dimensional reduction might get recovered as solitonic instanton particle states of the 5d MSYM theory. These correspond to D0 branes bound to the D4 branes. Their mass spectrum agrees with the KK momentum
\be
p_5= \frac{k}{R_5}
\ee
where $k$ is the instanton number and $R_5$ is the M-theory circle radius. For U(1) gauge group, the partition functions of
M5 on $T^6$ and of the dimensionally reduced D4 on $T^5$ have been computed explicitly. For U(1) gauge group, dimensional reduction does in fact truncates the KK modes and they are not recovered \cite{Dolan:2012wq}. However if we make an infinitesimal noncommutative deformation of the 5d MSYM theory, the KK modes are recovered as noncommutative instanton particles \cite{Bak:2012ct,Kim:2011mv}.

But the generalization to the nonabelian case is not so straightforward. The gauge coupling constant of 5d MSYM theory is dimensionful and is related to the circle radius $R_5$ by
\be
g^2_{\rm YM} = 4 \pi^2 R_5
\ee
and, hence, the theory is perturbatively nonrenormalizable. Even with the maximal number of supersymmetries in 5d, it turns out that 
the theory involves infinities beginning at six-loop order \cite{Bern:2012di}.  Therefore, the 5d MSYM theory as a definition of the M5 brane theory has some difficulties at the moment\footnote{However, localization computations in 5d SYM theory have produced  expected results of parallel M5 branes in flat Euclidean space. See for instance \cite{Minahan:2013jwa,Kallen:2012va,Kim:2012tr,Kim:2012ava,Kim:2012qf}.}. Nevertheless, in this paper we will find that this 5d MSYM  description of  M5/D4 is useful at least for BPS states with some remaining supersymmetries.

Below we shall be concerned with the Coulomb branch dynamics of $N$ D4 (M5) branes. The U(N) gauge symmetry is 
maximally 
broken down to U(1)$^{N-1}$  by the vev of the one of the scalar field\footnote{The overall U(1) gauge symmety will not be broken 
by the vev. In this paper we will not study S-duality for this overall U(1) gauge group, as this will require a separate treatment.}\label{tva}
\be
\langle \phi_6 \rangle = {\rm diag}\left[
v_1, v_2, \cdots, v_N\right]
\ee
where $v_i$ represents the location of $i$-th D4 (M5) brane in the 6th direction and we shall order 
\ben
v_1 < v_2  < \cdots  < v_N\label{ordering}
\een
without loss of generality. 

One candidate definition of the M5 brane theory  is the DLCQ ${\cal N}$=8 quantum mechanics of $k$ instantons, which may be 
used to compute  physical quantities (including net non-BPS contributions) within the $k$ instanton sector. The DLCQ limit 
of the $k$ D0 branes is described by 
the ${\cal N}$=8 quantum mechanics over the moduli space of  $k$ instantons whose metric can be obtained by the ADHM 
construction of the 5d MSYM theory. Our ${\cal N}$=8 quantum mechanics also involves a potential of the form
\be
  V= g_{rs} G^r G^s
\ee
where $G^r$ is the triholomorphic Killing vector whose form is determined by the vev of the scalar field.
The ${\cal N}$=8 supersymmetric completion is uniquely fixed by the moduli space metric $g_{rs}$ and the triholomorphic 
Killing vector $G^r$. Also in the DLCQ limit, this ${\cal N}$=8 quantum mechanics becomes exact, which is argued
in \cite{Lambert:2012qy,Bak:2013bba}. Using this DLCQ description, we have computed the 1/4-BPS index partition functions of  one ($k$=1) dyonic 
instanton \cite{Bak:2013bba} and found that the result agrees with that from the 5d MSYM theory based on the localization method \cite{Kim:2011mv}.  
Thus the  ${\cal N}$=8 quantum mechanics may be used to deal with the KK sector of M5 brane compactified on the M-theory circle
if we accept the M5/D4 correspondence. However a direct check of the validity of the proposal for the nonabelian case is not possible since
we do not know any direct formulation of the nonabelian M5 brane theory. 

An indirect test of the M5/D4 correspondence is the test of duality in 5d MSYM compactified on a circle. If we compactify $N$ parallel M5 branes on a two-torus with radii $R_4$ and $R_5$, we have a large diffeomorphism group SL($2,\mb{Z}$) that acts on the coordinates $x^4$ and $x^5$ of this two-torus. If we dimensionally reduce along $x^5$ we get a 5d MSYM on a circle with radius $R_4$, Yang-Mills coupling constant $g_{YM}^2=4\pi^2 R_5$ and gauge group U(N). Let us refer to this as theory $\A$. If we instead dimensionally reduce the M5 brane system along $x^4$ we get 5d MSYM theory on a circle with radius $R_5$ and Yang-Mills coupling constant $g_{YM}'^2=4\pi^2 R_4$ and again the gauge group is U(N) (See \cite{Tachikawa:2011ch} for furher details on how corresponding Lie algebras are transformed under this duality.). Let us refer to this as theory $\B$. Now if the M5/D4 correspondence is correct, it would not matter along which circle we dimensionally reduce. Both theory $\A$ and theory $\B$ would be dual to the same M5 brane theory, and so they would also be dual to each other.

As a check of this duality, we will show how certain 1/4-BPS states are mapped into each other under duality. In theory $\A$ we will consider 1/4-BPS dyonic instanton states. We will be mostly interested in the 1/4 BPS states that are associated with the singly connected maximal string F1 (06) from D4$_1$ to D4$_N$ along the 6th direction. (The digits in parentheses, here (06), will represent the worldvolume directions of branes or strings.) In a more general situation we have a singly connected string  from D4$_i$ to D4$_j$ with $i < j$, whose 1/4-BPS index works as a basic building block of the general  1/4-BPS multi-particle index of  dyonic instantons (see Eq.~(\ref{general})). In the decompactification limit $R_4 \rightarrow \infty$,
the 1/4 BPS index of a singly connected string has been computed in \cite{Kim:2011mv} from the 5d MSYM theory based on the
localization method. For definiteness, let us consider the single-particle 1/4-BPS index which corresponds to a singly connected maximal string for U(N) gauge group. With the compact
$x^4$ direction, we may also consider a D2 (046) which will have the finite mass
\be
M_{D2}= \frac{1}{2\pi g_s l_s }  R_4\, v =\frac{R_4}{ R_5}  \frac{v}{2\pi}
\ee
where $v$ is the difference of the vev's of the two associated D4 branes. The number of such D2 branes cannot be fixed due to the finite mass. However in the decompactification limit, the mass becomes infinite, and therefore we can put the number of D2 branes to be zero consistenly with the dynamics in the decompactification limit. We may also compare the mass of D2 with the mass of the F1 (06) connecting the same two D4 branes. This will have the mass
\be
M_{F1}= \frac{v}{2\pi}
\ee
and we see that we need $R_4\gg R_5$ in order for the D2 to be much heavier than the F1.

On the other hand, if $R_4\ll R_5$, then no matter how large we take $R_4$, we will always find that D2 branes are lighter than F1 strings, and then we can no longer consistently put the number of D2 to be zero. One must now consider more general 1/4-BPS states of the  singly connected  maximal string, which  involve relative charges of D2's
and $k\, (>0)$ D0's at the same time. The dynamics of these generalized 1/4 BPS states has not  been fully understood up to now and our study  below gives some prediction of their multiplet structures with nontrivial dependence on the relative D2 charges.

In theory $\A$ we will consider a dyonic instanton configuration which involves D4 (01234), D0 (0) and F1 (06). But as we argued above, if $x^4$ is compactified, we must to this system also add an unspecified number of D2 (046)\footnote{ 
The system involving additional D2's  is different from the 
1/4 BPS supertubes stretched between 
D4 branes
  \cite{Bak:2002ke}.   There is no  distinction between dyonic instantons and supertubes 
since both are 1/4 BPS and involve the same branes and strings.
The supertubes do not carry any D2 charges and, instead, it may carry a 
dipolar D2 brane charges which are not associated with any global symmetries.}\label{tre}. 

To see that theory $\A$ is dual to theory $\B$ (without resorting to the M5 brane theory), we first perform T-duality along $x^4$ which maps D4 into D3 of IIB string theory. We then make S-duality. We finally perform T-duality again, now along $x^5$ to get the D4 brane of theory $\B$. We will describe these duality maps in more detail in Section \ref{TST}. By TST duality the brane configuration turns into one with D4 (01235), W (05) and D2 (056) where $W$ represents a BPS-wave carrying momentum along the 5th circle direction. The decompactification limit $R_4 \rightarrow \infty$ of theory $\A$ becomes the strong coupling limit of theory $\B$ where the M-theory circle is along $x^4$. In theory $\B$,
we describe the system from the viewpoint of D2's which correspond
 to monopole strings wrapped around the 5th circle direction. To this system we also find the added F1 (06) which are TST duals of the added D2 (046). The low-energy dynamics of the monopole strings is governed by a 2d (4,4) nonlinear sigma model whose target space is given by the moduli space of  monopoles. 
In theory $\B$ the  singly connected  maximal D2 (056) (which is the TST dual of the singly connected maximal F1 (06)), the $4(N-1)$ dimensional monopole 
moduli space is described by the Lee-Weinberg-Yi (LWY) metric whose explicit form is known explicitly \cite{Lee:1996kz}. In particular for 
the U(3) gauge group,  the relative moduli space corresponds to the Taub-NUT (TN) space which we shall discuss in detail below.

This (4,4) nonlinear sigma model description becomes precise in the DLCQ limit of the 5th circle direction together with the weak coupling limit $R_5 \gg R_4$. 
The monopole moduli space involves  U(1)$^{N-1}$ isometry and the corresponding Noether charges are interpreted as electric charges of F1's 
connecting from D4$_i$ to D4$_{i+1}$ ($i=1,2, \dots, N-1$). The charge of the overall part will be denoted by $Q_{overall}$ while their relative 
charges by $Q_{relative}^{(m)}$ ($m=1,2, \dots, N-2$). Both the left-moving and right-moving Hamiltonians of the 2d sigma model commute with these electric charges, which means that we may refine the elliptic genus by introducing chemical potentials for these charges. We thus defined the refined elliptic genus
\be
\h Z(q,y, x) ={\rm tr}\, \, (-1)^{F_L+F_R} q^{L_0 -\frac{c}{24}} \bar{q}^{\bar{L}_0 -\frac{c}{24}} y^{J_L}\prod^{N-2}_{m=1} x_m^{Q_{relative}^{(m)}}\label{eg}
\ee
Here $J_L$ is a particular combination of the fermionic R-charge generators of the (4,4) sigma model that has an interpretation of the 
(2,2) supersymmetries. We may further refine by inserting $x_{ tot}^{Q_{overall}}$ inside the trace. But the elliptic genus will be independent of $x_{tot}$ because states with nonvanishing $Q_{overall}$ give no net contribution to the elliptic genus.

By the projection to zero relative electric charges, we may hope to compare our result with the 1/4-BPS index of  dyonic instantons in  \cite{Kim:2011mv}. In the decompactification limit, we may consistently put number of added D2 to be zero, which on the TST dual side corresponds to projecting down to the zero-charge sector of the elliptic genus. However, the decompactification limit of theory $\A$ is the strong coupling limit of theory $\B$. But our moduli space approximation of monopole string requires weak coupling. Nevertheless, we will successfully find a match with the result in \cite{Kim:2011mv}. The reason for this, is that our elliptic genus is really independent of the coupling constant $R_4$ of theory $\B$ so we can reliably compute it at weak coupling and then make the comparison with \cite{Kim:2011mv} by going to strong coupling. 

For  U(3) gauge group, our monopole moduli space has the form  \cite{Lee:1996if,Gauntlett:1996cw}
\be
\R^3 \times \frac{\R^1 \times M_{\rm TN}}{\mb{Z}}\label{Zid}
\ee   
The (refined) elliptic genus of the overall part, $\R^3 \times \R^1$,  is denoted as $Z_{com} (q,y)$, which is the same as the elliptic genus of $\R^4$ and is independent of $x_{\rm tot}$ under the 
refinement. The projection of the refined elliptic genus on TN space to the zero charge sector 
\be
Z^{\B}_0(q,y)=Z_{\rm com}(q,y)\oint_{x=0} \frac{dx}{2\pi i x}\h Z_{\rm TN} (q,y,x)
\ee     
where we attach the letter $\B$ to indicate this is computed in theory $\B$, is shown to agree with the 1/4-BPS index which was obtained for the dyonic instanton \cite{Kim:2011mv} in theory $\A$ in the decompactification limit $R_4\rightarrow \infty$. 

For the U(2) gauge group, 
the elliptic genus of the Atiyah-Hitchin (AH) space \cite{Atiyah:1985dv} is relevant for the dynamics of  two identical monopole strings.     
With an appropriate refinement, we shall verify the agreement of the two sides. 
For the case of U(N) with $N>3$, we shall compute the zero charge projection of the elliptic genus of the LWY space
by turning on potential related to the vev of another scalar field, which allows a localization in the moduli space, 
and again find an agreement 
of the two 
sides confirming the TST duality.

The charged 1/4-BPS sector of the monopole strings is of interest. The quantum mechanics of the zero mode part does not have this 1/4-BPS generalization. In addition, the overall part, even including its oscillator contribution, does not receive a net contribution from the charged sector.
Hence the generalization purely comes from the oscillator modes of the relative part of the 2d sigma model.  Hence
the relative F1's (06) are coming from the oscillator contribution, where the conventional rigid string interpretation of F1 
is broken down. This is  something that has a genuinely 2d character in the sense that it is not found in 
the corresponding moduli space dynamics of monopoles. In the TST dual side, the relative F1's correspond to the relative D2's and the corresponding 1/4-BPS generalization requires the presence of nonzero number of D0's. Further understanding in this direction will be of interest.

\section{Review of the 1/4-BPS dyonic instanton index}\label{beta}
In this section we will review the result that was obtained in \cite{Kim:2011mv}. But we will also make a small note on a refinement of the result that they presented by including the dependence of one chemical potential that we denote by $y$.

In the Coulomb branch of theory $\A$ with vev's $v_i$ ($i=1,2,\cdots,N$) ordered as described in the introduction, we break the gauge group maximally down to U(1)$^{N-1}$ (times the center of mass U(1)). We have sectors labeled by the instanton number $k=1,2,...$. In each instanton sector an index is defined as\footnote{We use capital $X_i$ for the $U(1)$'s here to emphasise that these are actually different from the $x_m$ that were introduced in the elliptic genus in Eq (\ref{eg}).}\label{fyra}
\bea
I_{k,N}(X_i,y) &=& \tr_k \((-1)^F e^{-\beta H} X_1^{\Pi_1}\cdots X_N^{\Pi_N} y^J\)
\eea
Here $H$ denotes the Hamiltonian of the D4-D0 system, $\Pi_{i}$ is the U(1)$_i$ gauge group generator, with a corresponding chemical potential $X_i = e^{-\mu_i}$, and $J$ is one of the Cartan generators of the little symmetry group $SO(4)\times SO(4) \subset SO(1,4)\times SO(5)$ of a massive object in presence of a vev, with corresponding chemical potential $y=e^{2\pi i z}$. To match with the notation in reference \cite{Kim:2011mv} we shall take $2\pi z = \gamma_2$ and $J=-2J_{2L}$. We view the other two chemical potentials $\gamma_1$ and $\gamma_R$ in \cite{Kim:2011mv} as regulators that we take towards zero. 

As usual, the index does not directly depend on $\beta$ which we may take towards zero. This localizes the path integral to a set of saddle points, which enables explicit computation of the index. This has been done in \cite{Kim:2011mv} where it was found that the index in the sector $(k,N)$ is a sum of contributions $I_{\{Y_1,...,Y_N\}}$ where $\{Y_1,....,Y_N\}$ denotes a set of $N$ Young diagrams, some of which can be empty, and where $k$ boxes are distributed over all these Young diagrams. The contribution from one set of such Young diagrams is given by\footnote{The same formula is obtained in \cite{Bruzzo:2002xf} for mass-deformed 4d N=2 SYM, but with sinh linearized.}\label{fem}
\ben
I_{\{Y_i,...,Y_N\}} &=& \prod_{i\in {\cal Y}} \prod_{j=1}^N \prod_{s\in Y_i} \frac{\sinh \frac{E_{ij}(s)-i(\gamma_2+\gamma_R)}{2}  \sinh \frac{E_{ij}(s)+i(\gamma_2-\gamma_R)}{2}} {\sinh \frac{E_{ij}(s)}{2} \sinh \frac{E_{ij}(s)-2i\gamma_R}{2}}\label{general}
\een
and to get the index $I_{k,N}$ we shall sum over all such sets. In this formula, the index $i$ is taken from a set ${\cal Y} \subset \{1,...,N\}$ which corresponds to Young diagrams that are not empty. We can then pick a box $s=(m,n)\in Y_i$ at row $m$ and column $n$, and to it assign
\bea
E_{ij}(s) &=& \mu_i - \mu_j + i(\gamma_1-\gamma_R) h_i(s)+i (\gamma_1+\gamma_R)(v_j(s)+1)
\eea
where
\bea
h_i(s) &=& \nu_{im} - n\cr
v_j(s) &=& \nu'_{jn} - m
\eea
and $\nu_{im}$ denotes the number of boxes in row $m$ in $Y_i$ and $\nu'_{jn}$ denotes number of boxes in column $n$ in $Y_j$. We define $\nu_{im}=0$ if there are no boxes at that row, or if the whole Young diagram $Y_i$ is empty.

The multi-particle index is given by
\bea
I_N &=& \sum_{k=0}^{\infty} I_{k,N} q^k
\eea
where we put $I_{0,N} = 1$. The expansion parameter is given by $q^k=e^{-S(k)}$ where $S(k)$ denotes the Euclidean 5d MSYM classical action evaluated at instanton number $k$,
\bea
S(k) = \frac{1}{g_{YM}^2} \int_0^{2\pi\beta} dt \int d^4 x \frac{1}{4} \tr(F_{ij} F_{ij}) = 2\pi\beta\frac{k}{R_5}
\eea
If we include a graviphoton (which is the up-lift to 5d of the theta parameter in 4d SYM) in the action, this will complexify the Euclidean action. We may then define
\ben
q &=& \exp 2\pi i \tau\label{q}
\een
where
\bea
\tau &=& i\frac{\beta}{R_5}
\eea

In theory $\B$ after TST-duality, we have monopole strings. The energy of a BPS wave on the monopole string is fixed by the BPS equation that we derive from the 2d (4,4) sigma model. This BPS equation is given by $H=P$ where $H$ is the sigma model Hamiltonian, and $P$ is the momentum along the monopole string. By the fact that the monopole string is circle-compactified with radius $R_5$, it follows that, for BPS states (from the 2d sigma model viewpoint), $H=P=k/R_5$. The expansion parameter in the elliptic genus will again be given by (\ref{q}) with the same $\tau = i\frac{\beta}{R_5}$. In a more general situation we can in both theory $\A$ and theory $\B$ also have a real part of the complex parameter $\tau$. From the M5 brane viewpoint this is the $\tau$-parameter of the two-torus spanned by Euclidean time $x^0$ and $x^5$. Note that the S-duality we consider acts on the two-torus which is spanned by $x^4$ and $x^5$ and therefore S-duality does not act on the above $\tau$-parameter. 

From the multi-particle index $I_N$ we can extract the single particle index $z_{sp}$ (which is expected to correspond to the index of monopole strings, in a way that we will clarify a bit further below) from the plethystic exponential
\bea
I(q,\gamma) &=& \exp \sum_{n=1}^{\infty} \frac{1}{n} z_{sp}(q^n,n\mu,n\gamma)
\eea
Following \cite{Kim:2011mv}, we factor out the divergent factor $I_{com}(\gamma)$ from $z_{sp}$ and define
\bea
z_{sp}(q,\mu,\gamma) &=& I_{com}(\gamma)z'_{sp}(q,\mu,\gamma)
\eea
For generic $N$ we define $X_{ij}=e^{-(\mu_i-\mu_j)}$ which is the chemical potential of an M2 brane (that is, an F1 (06) in the theory $\A$ description and a D2 (056) in the theory $\B$ description) stretching between M5 branes (D4 branes) $i$ and $j$. If we have $n$ M2 branes stretched from M5$_i$ to M5$_j$ this will come with chemical potential $X_{ij}^n$. In other words, by expanding $z'_{sp}$ in powers of $X_{ij}$ we extract the contribution from $n$ M2 branes between M5$_i$ and M5$_j$ by reading off the coefficient of $X_{ij}^n$. 

In the Appendix \ref{generalapp} we extract the following single particle indices from the general formula (\ref{general}),
\ben
z'_{N=2,n=0}(q,y) &=& 2q+\Ordo(q^2)\cr
z'_{N=2,n=1}(q,y) &=& 1 + \(4-2(y+y^{-1})\)q + \(18-10(y+y^{-1})+y^2+y^{-2}\) q^2 + \Ordo(q^3)\cr
z'_{N=2,n=2}(q,y) &=& \(8-4(y+y^{-1})\) q + \(112 -72(y+y^{-1})+16(y^2+y^{-2})\) q^2 + \Ordo(q^3)\cr
z'_{N=2,n\geq 2}(q,y) &=& 0+2n(2-y-y^{-1})q+\Ordo(q^2)\cr
z'_{N=3,n=1}(q,y) &=& 1+ \(10-6(y+y^{-1})+y^2+y^{-2}\) q + \Ordo(q^2)\label{orderq2}
\een
The series expansion for $z'_{N=2,n=1}$ matches with what one gets when one expands out the following proposed closed formula \cite{Kim:2011mv} 
\ben
z_{N=2,n=1} = -\frac{\theta_1(q,yu)\theta_1(q,yu^{-1})}{\theta_1(q,u)^2}= I_{com} z'_{N=2,n=1}\label{su2} 
\een
We summarize the theta functions and their modular transformations in the Appendix \ref{theta}. We also can see that 
\ben
z'_{N=N,n=1}(q,y) &=& z'_{N=2,n=1}(q,y)\(Z^{\A}_0(q,y)\)^{N-2}\label{Z0}
\een
where
\ben
Z^{\A}_0(q,y) &=& 1+ \frac{(1-y)^4}{y^2} q + \Ordo(q^2)\label{Z0expansion}
\een
Here we have only verified (\ref{su2}) up to order $q^2$ and (\ref{Z0expansion}) up to order $q$ and for the case when $N=3$. But these series expansions were presented at $y=-1$ to higher orders in $q$ in \cite{Kim:2011mv},
\ben
z'_{N=2,n=1}(q,-1) &=& 1 + 8 q + 40 q^2 + 160 q^3 +\cdots + 188784 q^{10} + \Ordo(q^{11})\cr
z'_{N=2,n=2}(q,-1) &=& 0 + 16q + 288 q^2 + 2880 q^3 + \cdots + 125280 q^5 + \Ordo(q^6)\cr
z'_{N=3,n=1}(q,-1) &=& 1 + 24 q + 246 q^2 + 264 q^3 + 2016 q^3 + \cdots+290976q^6+\Ordo(q^7)\cr
z'_{N=4,n=1}(q,-1) &=& 1 + 40q + 774q^2 + 8992q^3+82344q^4+\Ordo(q^5)\cr
z'_{N=5,n=1}(q,-1) &=& 1+56q+1480q^2+25184q^3+317288q^4+\Ordo(q^5)\label{high1}
\een
In this reference the closed formulas (\ref{su2}) and (\ref{Z0}) were verified up to these orders at $y=-1$, and a series expansion for $Z^{\A}_0(q,-1)$ was extracted as
\ben
Z^{\A}_0(q,-1) &=& 1+16 q + 96 q^2 + 448 q^3 + \cdots+ 18048 q^6 + \Ordo(q^7)\label{high2}
\een

\section{Some basics of the elliptic genus}
Before going into detailed computations of elliptic genera, we recall some basics \cite{Witten:1993jg}. 
The elliptic genus of a (2,2) superconformal\footnote{This condition can be relaxed and we can still define an elliptic genus, but we will not need that here.}\label{sex} 2d sigma model can be defined as
\bea
Z(q,y) &=& \tr (-1)^{F_L+F_R} y^{J_L} q^{H_L} \bar{q}^{H_R} 
\eea
where we define the left-moving and right-moving Hamiltonians as
\bea
H_L = L_0 - \frac{c}{24} = \frac{H+P}{2}\cr
H_R = \bar{L}_0 - \frac{c}{24} = \frac{H-P}{2}
\eea
where $H$ and $P$ are the time translation and space translation generators of the 2d sigma model. The trace is over all states in the Hilbert space. Here $J_L$ is acting non-trivially only on the left-moving sector. Then due to the insertion of $(-1)^{F_R}$ all states with non-vanishing $H_R$ mutually cancel out in the elliptic genus, so that only states which saturate the BPS bound 
\ben
H &=& P\label{bound1}
\een
contribute. We may think on $P$ as a central charge which appears in the superalgebra. This means that although we insert $\bar{q}^{H_R}$ into the trace, the elliptic genus will only depend holomorphically on $q$. The BPS bound of $H$ can actually be higher than $P$ if there are other central charges present in the 2d sigma model supersymmetry algebra (see footnote \ref{tio}). So the bound (\ref{bound1}) may be lower than the BPS bound. We therefore like to avoid referring to states saturating the bound (\ref{bound1}) as BPS states. We will instead refer to states which saturate the bound (\ref{bound1}) as left-moving BPS states. 

What we said so far applies only if there is no continuum of states contributing to the elliptic genus. If there is a continuum part, then the trace shall be replaced by an integral over the energy and weighted by the density of states $\rho(E)$ at each energy level. It is now possible for the densities for bosonic and fermionic states to be different from each other, in which case we find contributions coming from states with a nonzero $H_R$. In other words, if there is a contribution coming from a continuum of states, then this will come as a non-holomorphic term in the elliptic genus, and the elliptic genus can be separated into two pieces, discrete plus coninuum. See for example \cite{Giveon:2014hfa} for more details on the continuum part.

\section{The elliptic genus for $\mb{R}^3 \times S_e^1$ sigma model}
Let us begin with U(2) gauge group and a single fundamental monopole string. The 2d sigma model that lives on this monopole string has (4,4) supersymmetry and the target space $\mb{R}^3\times S^1_e$, which is the same as the moduli space of a single fundamental SU(2) monopole. For a generic radius on the gauge circle $S^1_e$ (in relation to the radius $R_5$ of the monopole string), there are no left-moving states that carry non-vanishing momentum $Q_{overall}$ along $S^1_e$.\footnote{Existence of left-moving states with non-vanishing momentum $Q_{overall}$ is possible if we also have for example a winding number also equal to $Q_{overall}$ and if the gauge circle radius equals $R_5$. We will not consider such special situations in this paper, but will assume the radii are generic. In that case no winding strings will be left-moving.}\label{sju} Therefore the elliptic genus can equally well be computed with the target space being replaced with $\mb{R}^4$. The elliptic genus for 2d (2,2) sigma model with flat target space $\mb{C}=\mb{R}^2$ has been obtained in \cite{Witten:1993jg}. The result is 
\ben
Z(q,y,x) &=& \frac{\theta_1(q,yx)}{\theta_1(q,x)}\label{u1}
\een
Here $y$ is the chemical potential associated to the R-charge and $x$ is associated to the global U(1) symmetry of the target space $\mb{C}$ of the (2,2) sigma model. 
 
For our 2d (4,4) sigma model on $\mb{R}^4$ the result becomes
\ben
Z(q,y,x) &=& \frac{\theta_1(q,yx)\theta_1(q,yx^{-1})}{\theta_1(q,x)\theta_1(q,x^{-1})}\label{eg11}
\een
The target space $\mb{R}^4$ has $SO(4) = SU(2) \times SU(2)$ rotation symmetry. The SU(2) R-symmetry of the (4,4) sigma model is generated by the three Kahler forms $J^{I+}$. On $\mb{C}^2$ these can be realized as selfdual 't Hooft matrices $\eta^{I+}_{ij}$. The other commuting global U(1) symmetry must therefore be generated by $\eta^{I-}_{ij}$ which acts on both fermions and bosons. Thus the elliptic genus is defined as
\bea
Z(q,y,x) &=& \tr_{RR}(-1)^F y^{J^{3+}} x^{J^{3-}}q^{L_0 -\frac{c}{24}} {\bar{q}}^{\bar{L}_0 -\frac{c}{24}} 
\eea
We notice that the elliptic genus (\ref{eg11}) agrees with the corresponding index (\ref{su2}) in theory $\A$.

We also notice that (\ref{eg11}) satisfies the spectral flow equation for $\hat{c}=c/3=2$,
\ben
Z(q,y q^m,x) &=& q^{-m^2} y^{-2m} Z(q,y,x)\label{spfl}
\een
Here $c$ denotes the central charge, which for our (4,4) sigma model on 4d target space is given by $4\cdot 1 + 4 \cdot \frac{1}{2} = 6$ for four bosons and four fermions. We will return to this spectral flow equation in more detail in the next section. To check that (\ref{eg11}) satisfies (\ref{spfl}), it is enough to take $m=1$ and use Eq.~(\ref{periodb}).

We notice that two data points are particular simple. Namely at the two points $y=1$ and $y=-1$. Here the elliptic genus reduces to 
\bea
Z(q,1,x) &=& 1\cr
Z(q,-1,x) &=& \(\frac{\theta_1(q,-x)}{\theta_1(q,x)}\)^2
\eea
where we have noticed that $\theta_1(q,x^{-1}) = -\theta_1(q,x)$. At these data points, the  elliptic genus becomes the square of the elliptic genus for the (2,2) sigma model. We can thus be confident about the correctness of the result at these two data points. But we can then uniquely deduce the full elliptic genus for any $y$ just using these two data points and the fact that it shall satisfy the spectral flow equation.

\section{The elliptic genus for TN sigma model}
The elliptic genus on TN was obtained in \cite{Harvey:2014nha} and is given by 
\ben
Z(q,y;x) &=&  \frac{g^2}{\tau_2} \int_{\mb{C}} du d\bar{u} \frac{\theta_1(q,yxz)\theta_1(q,yx^{-1}z^{-1})}{\theta_1(q,xz)\theta_1(q,x^{-1}z^{-1})} e^{-\frac{g^2 \pi}{\tau_2} |u|^2}\label{eg1}
\een
where $z = e^{2\pi i u}$ and $q = e^{2\pi i \tau}$ where $\tau = \tau_1 + i\tau_2$ using our notations.  Here $g$ is a size parameter in the TN metric and $x$ is the chemical potential associated to the U(1) isometry of TN.

There is a contribution from a discrete set of winding monopole string states around this TN circle and which are BPS by a balancing angular momentum along the circle, as well as from discrete non-winding states. But to the elliptic genus there are also states in a continuum which contribute, which means that the elliptic genus contains a non-holomorphic term. 

For compact target manifolds, the elliptic genus reproduces the Euler characteristic $\chi$ and the Hirzebruch signature $\sigma$ in two different limits,
\bea
Z(q,1;1) &=& \chi\cr
\lim_{q\rightarrow 0} Z(q,-1;1) &=& \sigma
\eea
In this case the target manifold is TN which is noncompact, and so one may expect some additional complications. What we find is that 
\bea
Z(q,1;1) &=& 1\cr
\lim_{q\rightarrow 0} Z(q,-1;x) &=& {\mbox{pole singularity as $x\rightarrow 1$}}
\eea
It is easy to see that the Euler  characteristic on TN \cite{Gibbons:1979xm} is reproduced this way from the elliptic genus \cite{Harvey:2014nha}. To extract the signature from the elliptic genus we shall take the limit $q\rightarrow 0$. This amounts to taking $\tau_2 \rightarrow \infty$. We then find a delta function $\delta(u,\bar{u})$ that picks out the elliptic genus on $\mb{R}^4$ from the elliptic genus on TN. The signature on TN is known to be zero \cite{Gibbons:1979xm}, but what we find is that the elliptic genus on TN has a pole singularity as $x$ approaches to $x=1$, which comes from the pole in the elliptic genus on $\mb{R}^4$ if we remove the chemical potential. But on the other hand, if we remove this chemical potential, then additional fermionic zero modes arises on $\mb{R}^4$ which puts the whole elliptic genus on $\mb{R}^4$ to be zero, which is what want to match with the known value of the signature of TN. So the pole shall be replaced by zero. The elliptic genus on $\mb{R}^4$ has a discontinuity as we turn off the chemical potential due to this fermionic zero mode that kicks in and puts the elliptic genus to zero, when we turn it off. In other words lim$_{x\rightarrow 1} Z(0,y;x) \neq Z(0,y;1) = 0$.

Here our main interest in the TN elliptic genus will be in its discrete part which is holomorphic. In particular we will be interested in its zero charge sector. This was extracted from the full elliptic genus (\ref{eg1}) in \cite{Harvey:2014nha}. The result that was found can be expressed as
\ben
Z_0(q,y) &=& \oint_C \frac{dx}{2\pi i x} \frac{\theta_1(q,yx)\theta_1(q,yx^{-1})}{\theta_1(q,x)\theta_1(q,x^{-1})}\label{eg10}
\een
where the integration contour $C$ is defined as $|x|=r$ where $|q|<r<1$. We will give an other argument for this result for the zero charge sector in section \ref{zerocharge}.

\section{The elliptic genus for AH sigma model}
There is an argument in  \cite{Harvey:2014nha} that says that we can have a contribution to the elliptic genus from scattering states only when the non-compact target space has a finte circle $S^1$ at infinity. Furthermore it is known that scattering states can lead to non-holomorphic terms in the elliptic genus. For AH we have SO(3) isometry and there is a U(1) embedded in SO(3) which corresponds to a circle direction in AH. However, this circle grows to infinite size at infinity. We therefore think that AH case is different from TN case. In the asymptotic region of AH we have locally essentially the same geometry as for TN. But the circle at infinity in the TN case is not mapped into an exact isometry of AH space. By this argument, we believe that there will be no continuum of states that contribute to the elliptic genus for the AH case. 

Based on this assumption, we would now like to obtain the elliptic genus of the 2d (4,4) AH sigma model. The Euler characteristic $\chi(AH)=2$ and the signature $\sigma(AH) =1$ of AH were obtained quite recently in \cite{Atiyah:2011fn}\footnote{We thank Nigel Hitchin for providing us with this reference.}. 
The separation of these numbers into a bulk contribution and contributions from the boundary were also obtained  \cite{Atiyah:2011fn}
\bea
\chi_{bulk}(AH) &=& 2\cr
\sigma_{bulk}(AH) &=& \frac{4}{3}
\eea
As was explained in \cite{Atiyah:2011fn}, the fall off at infinity is too fast for there to be any contributions from boundary integrals over local quantities to neither the Euler  characteristic nor the signature for AH as well as for TN. But for the signature there is a contribution from the eta invariant (which can not be expressed as a boundary integral over a local quantity) which for AH is $-\frac{1}{3}$ so that in total the signature of AH is $\sigma(AH) = 1$.  

The bulk contribution to the elliptic genus is given by \cite{Kawai:1993jk}
\ben
Z(q,y) &=& \int_{M_{2n}} \prod_{a=1}^{n} x_a \frac{\theta_1(q,y e^{2x_a})}{\theta_1(q,e^{2x_a})}\label{Kawai}
\een
where $x_a$ are the Chern roots associated to the curvature of $M_{2n}$. Applying this formula we find the bulk contribution
\bea
Z_{bulk}(q,y) &=& \frac{2}{3} \[\(\frac{\theta_2(q,y)}{\theta_2(q,1)}\)^2 + \(\frac{\theta_3(q,y)}{\theta_3(q,1)}\)^2 + \(\frac{\theta_4(q,y)}{\theta_4(q,1)}\)^2\]
\eea
which is $SL(2,\mb{Z})$ covariant. There is also a reason for this. Namely, in \cite{Kawai:1993jk}, it was shown that $SL(2,\mb{Z})$ covariance is automatic when the first Chern-class $x_1 + x_2$ is zero. Here we can confirm that this is the case by computing
\bea
\int_{AH} (x_1+x_2)^2 = \int_{AH} \(x_1^2 + x_2^2\) + 2\int_{AH} x_1 x_2 = -3 \cdot \frac{4}{3} + 2\cdot 2 = 0
\eea
where we note that the Euler  characteristic and the signature are given by
\bea
\sigma &=& -\frac{1}{3} \int \(x_1^2 + x_2^2\)\cr
\chi &=& \int x_1 x_2
\eea
We may check that the same argument goes through for the K3 elliptic genus which is $SL(2,\mb{Z})$ covariant. For K3 which is compact, we only have bulk contributions and $(\chi,\sigma) = (16,24)$. With this we get $\int_{K3} (x_1+x_2)^2 = - 3 \cdot 16 + 2 \cdot 24 = 0$.

In $Z_{bulk}$ for AH we find fractional coefficients, which indicate that something is missing. Clearly for $Z_{bulk}(0,-1)$, what is missing, is the eta invariant. To get the correct result for the elliptic genus at least at the points $y=1$ and $y=-1$ we shall use eq (\ref{Kawai}), but when we expand out the integrand, we shall everywhere replace the expression $-\int \frac{1}{3}\(x_1^2 + x_2^2\)$ with the topologically invariant expression $-\int\frac{1}{3}\(x_1^2 + x_2^2\) - \frac{1}{3}$ where we subtract the eta invariant. We have regularization problems when deriving eq (\ref{Kawai}) at all points except $y=1$ and $y=-1$ if we use path integral methods which keeps diffeomorphism invariance manifest. We find no regularization problems if we use the Hamiltonian quantization method, but in that case diffeomorphism invariance is not manifest, and it could get lost in the quantization procedure by a diffeomorphism anomaly if we include boundary contributions. We therefore shall trust eq (\ref{Kawai}) only at $y=1$ and $y=-1$. At these points we find 
\bea
Z(q,-1) &=& \frac{1}{2} \[\(\frac{\theta_3(q,y)}{\theta_3(q,1)}\)^2 + \(\frac{\theta_4(q,y)}{\theta_4(q,1)}\)^2\]\cr
Z(q,1) &=& 2
\eea
From this we find that there appears to be a unique $SL(2,\mb{Z})$ modular covariant solution which satisfies spectral flow and which interpolates between these boundary data. This solution is given by
\bea
 Z(q,y;x) 
& = &  \frac{1}{2} \left[ \ \  \frac{\theta_1(q,yx)\theta_1(q,yx^{-1})}{\theta_1(q,x)\theta_1(q,x^{-1})} + \frac{\theta_2(q,yx)\theta_2(q,yx^{-1})}{\theta_2(q,x)\theta_2(q,x^{-1})} \right.\cr
&& \ \  +\left. \frac{\theta_3(q,yx)\theta_3(q,yx^{-1})}{\theta_3(q,x)\theta_3(q,x^{-1})}+
\frac{\theta_4(q,yx)\theta_4(q,yx^{-1})}{\theta_4(q,x)\theta_4(q,x^{-1})}\ \ \right]
\eea
The refinement using the chemical potential $x$ is possible thanks to the SO(3) isometry of AH space, so we can pick a chemical potential and associate this to the Cartan generator of this SO(3). The form of this refinement is uniquely determined by spectral flow as we will see in the next section. We are forced to consider a refined version of the elliptic genus since the first term is singular if we drop the chemical potential, or in other words the limit $x\rightarrow 1$ is divergent. We have explained that this will have a discontinuity at $x=1$ where this term shall be zero due to additional fermionic zero modes at this point. With these considerations, this elliptic genus correctly reproduces the Euler characteristic $Z(q,1;1) = 2$ and the signature $Z(q\rightarrow 0,-1;1) = 1$, and for generic $q$ at $x=1$ it reproduces the boundary data $Z(q,-1;1) = Z(q,-1)$ due to these fermionic zero modes.

To reach our result for the elliptic genus on AH we have made some guess of what the result could be, as well as we have applied (\ref{Kawai}) on a case where it is not directly applicable. Our result should therefore be viewed as a conjecture. 

The Euler characteristic of AH is $\chi = 2$ and it corresponds to the Euler  characteristic of the Bolt, which is $\mb{R}^2$ fibered over $S^2$. For fiber bundles we have that the Euler characteristic of the bundle is the product $\chi = \chi($fiber$)\chi($base manifold$)$. We compute $\chi(\mb{R}^2)$ as the contribution of the bulk plus a contribution from a boundary circle at infinity. This amounts to saying that $\chi(\mb{R}^2) = \chi($disk$) =1$. We also have $\chi(S^2) = 2$. For the Bolt we have two harmonic forms. One is the volume form of $S^2$ and the other is the Hodge dual of this. But only one linear combination of these is normalizable on AH \cite{Sen:1994yi}. We like to count only normalizable states. This means that we like to subtract $1$ from the Euler  characteristic. To achieve this counting in the elliptic genus, we must everywhere replace $\int x_1 x_2$ with $\(\int x_1 x_2\) - 1$ in the expansion using eq (\ref{Kawai}), as well as substituting $-\int \frac{1}{3}\(x_1^2 + x_2^2\)$ with the topologically invariant expression $\(-\int \frac{1}{3}\(x_1^2 + x_2^2\)\) - \frac{1}{3}$ at the same time. If we do this, then eq (\ref{Kawai}) gives the result
\bea
Z(q,-1) &=& \frac{1}{2} \[\(\frac{\theta_3(q,y)}{\theta_3(q,1)}\)^2 + \(\frac{\theta_4(q,y)}{\theta_4(q,1)}\)^2\]\cr
Z(q,1) &=& 1
\eea
If we now use spectral flow with these two boundary data at $y=1$ and $y=-1$, we obtain
\ben
Z(q,y) &=& \frac{1}{2} \[\(\frac{\theta_3(q,y)}{\theta_3(q,1)}\)^2 + \(\frac{\theta_4(q,y)}{\theta_4(q,1)}\)^2\]\label{AH2}
\een
and this expression has integer coefficients, suggesting that this might be a correct result. Again there will be a unique refined version of this where we restore the chemical potential $x$.

What we really would like to do, is to refine the AH elliptic genus with respect to the discrete $\mb{Z}_2$ isometry of AH space. We will discuss this further in section \ref{section7}. But we have no idea how to do that. Instead we do what we can, and consider another refinement with respect to the Cartan of the continuous $SO(3)$ isometry, to which we associate the chemical potential $\mu$ and we define $x = e^{2\pi i \mu}$. Let us introduce the quantities
\ben
f_a(q,y,x) &=&  \frac{\theta_a(q,yx)\theta_a(q,y^{-1}x)}{\theta_a(q,x)^2}
\een
for $a=1,2,3,4$ labeling the four different theta functions. In eq (\ref{reg}) we will obtain the refined version of the elliptic genus (\ref{AH2}) as
\ben
\h Z(q,y,x) &=& \frac{1}{2} \[f_3(q,y,x) + f_4(q,y,x)\] \label{AH2ref}
\een
Let us now note the following theta function identities
\bea
\theta_4(q,y) &=& i q^{\frac{1}{8}} y^{-\frac{1}{2}} \theta_1(q,q^{\frac{1}{2}} y)\cr
&=& -i q^{\frac{1}{8}} y^{\frac{1}{2}} \theta_1(q,q^{-\frac{1}{2}} y)
\eea
as well as
\bea
\theta_3(q,y) &=& \theta_4(q,-y)
\eea
Using these we find 
\bea
f_4(q,y,x) &=& f_1(q,y,q^{-\frac{1}{2}} x)\cr
&=& f_1(q,y,q^{\frac{1}{2}} x)\cr
f_3(q,y,x) &=& f_4(q,y,-x)
\eea
The theta function $\theta_4(q,x)$ has zeroes at $x = q^{n+\frac{1}{2}}$ for $n\in \mb{Z}$. We can therefore Laurent expand $f_4$ around $q=0$ and such a Laurent expansion will be valid in the disk $0\leq |q|<|x|^2$. Hence we shall consider a contour $C_{\frac{1}{2}}$ defined as $|x| = x_0$ such that $|q|^{\frac{1}{2}} < x_0$.

The function 
\bea
\varphi^F(q,y) &=& \oint_{C_0} \frac{dx}{2\pi i x} f_1(q,y,x) 
\eea
was studied in \cite{Harvey:2014nha} where the contour $C_0$ was defined to lie in the annulus $|q|<|x|<1$. 

This translates into the contour $C_{-\frac{1}{2}}$ inside the annulus $|q|^{\frac{1}{2}} < |x'| < |q|^{-\frac{1}{2}}$ for the variable $x' = q^{-\frac{1}{2}} x$. It translates into another contour $C_{\frac{1}{2}}$ in the annulus $|q|^{\frac{3}{2}} < |x''| < |q|^{\frac{1}{2}}$ for the variable $x'' = q^{\frac{1}{2}} x$. 

We conclude that we also have
\bea
\varphi^F(q,y) &=& \oint_{C_{\pm\frac{1}{2}}} \frac{dx}{2\pi i x} \frac{1}{2} \[f_3(q,y,x) + f_4(q,y,x)\]
\eea
and that either sign gives the same answer. 

The contour integral over $x$ picks out the contribution from states with zero $SO(3)$ charge. 

We will then make a very convincing numerical check in eq (\ref{charged}) which, as we explain further in section \ref{section7}, shows the following correspondence: states with zero $SO(3)$ charge are odd under the $\mb{Z}_2$ isometry, and states which have a nonzero $SO(3)$ charge are even under $\mb{Z}_2$. Alternatively, if one can justify this correspondence by some independent means, then this would amount to a check of S-duality. One can directly confirm that this correspondence is valid for Sen's ground state two-form harmonic wave function on AH \cite{Sen:1994yi}. Here our result shows that this correspondence is true for all BPS states which contribute to the elliptic genus, once we accept the S-duality hypothesis. A direct check of this correspondence might require an explicit construction of these wave functions.

In the next section we will show in more detail how we reached our result using spectral flow. We also show that the latter elliptic genus (\ref{AH2}) or its refinement (\ref{AH2ref}) is not fully $SL(2,\mb{Z})$ covariant. Clearly the lack of full $SL(2,\mb{Z})$ covariance must have come about in the process of removing the non-normalizable mode, or in other words, when we subtract $1$ from the Euler characteristic by hand.

\subsection{Derivation of elliptic genus using spectral flow}
We assume that the 2d sigma model on AH is exactly superconformal. We then use the (2,2) superconformal part of the original (4,4) superconformal symmetry of the sigma model. One can show that the spectral flow \cite{Schwimmer:1986mf}
\bea
L_n &\rightarrow& L_n + \alpha J_n + \frac{\alpha^2}{2}\hat{c}\delta_{n,0} \cr
J_n &\rightarrow& J_n + \alpha \hat{c} \delta_{n,0} \cr
G^\pm_n &\rightarrow& G^\pm_{n\pm \alpha}
\eea
is the inner automorphism of the superconformal theory,
where $\alpha$ is an arbitrary real parameter. 

Since $J_0$, which is identified with $J_L$, is 
integral or half-integral quantized depending on $\hat{c}$,  one has\footnote{We define $q=e^{2\pi i \tau}$ and $y = e^{2\pi i z}$. When it is more convenient, we will view the elliptic genus as a function of $\tau$ and $z$ instead of $q$ and $y$.}\label{nio}
\be
Z(\tau, z+n)=(-1)^{\hat{c}n} Z(\tau,z)
\ee
for $n \in \mathbb{Z}$. Using the spectral flow and taking $\alpha=m \in \mathbb{Z}$,
the RR sector remains to be RR after the spectral flow and it is straightforward to show that
\be
Z(\tau, z+m \tau+n)= (-1)^{\hat{c}(m+n)} e^{-\pi i \hat{c} ( m^2 \tau + 2 m z)}\, Z(\tau, z)
\ee
The most general solution of this constraint is given by an arbitrary linear combination of
\be
\theta_{(l)} (\tau, z)= \sum_{n\in \mathbb{Z}} q^{\frac{\hat{c}}{2}(n + \frac{l}{\hat{c}})^2} y^{\hat{c} n + l} 
\ee
whose  coefficients are functions of only $\tau$ and 
\be
 l= -\frac{\hat{c}}{2},-\frac{\hat{c}}{2}+1, \cdots, \frac{\hat{c}}{2}-1
\ee
 Namely the solution is
\be
Z(\tau, z)=\sum^{\frac{\hat{c}}{2}-1}_{l= -\frac{\hat{c}}{2}} h_l(\tau) \,  \theta_{(l)} (\tau, z)
\label{ega}
\ee
The consistency with the spectral flow requires that
\be
h_l(\tau) =\sum^\infty_{k=0} C_{kl} \, q^{k+D(l)}
=\sum^\infty_{k=0} C\Big(k-\frac{l^2}{2\hat{c}}\Big) \, q^{k-\frac{l^2}{2\hat{c}}}
\ee
To show this, we note that
\be
 Z(\tau, z)=\sum_{l,k,n} C_{kl} q^{k+D(l)+\frac{\hat{c}}{2}(n + \frac{l}{\hat{c}})^2} y^{\hat{c} n + l} 
\ee
To be consistent with the RR spectra, $D(l)$ should be chosen as
\be
D(l)= -\frac{l^2}{2\hat{c}}
\ee
and then the consistency with the spectral flow requires
\be
C_{kl}= C\Big(k-\frac{l^2}{2\hat{c}}\Big)=C\Big(k-\frac{l^2}{2\hat{c}}
+\frac{\hat{c}}{2}\Big(n + \frac{l}{\hat{c}}\Big)^2 - \frac{(\hat{c} n + l)^2}{2\hat{c}}
\Big)
\ee
since the combination $L_0 - \frac{1}{2\hat{c}}J_0^2$ is invariant under the spectral flow.

For the elliptic genus of the (4,4) sigma model, $\hat{c}=2$ and hence (\ref{ega}) involves
two independent functions of $\tau$, $h_0(\tau), h_{-1}({\tau})$. We also have the boundary data
\ben
Z(\tau,0) &=& 1\cr
Z\(\tau, 1/2\) &=& \frac{1}{2} \left[
\left(\frac{\theta_3 (\tau,1/2)}{\theta_3 (\tau,0)}\right)^2 + \left(\frac{\theta_4 (\tau,1/2)}{\theta_4 (\tau,0)}\right)^2
\right] \cr
&=& 1+ 32 q + 256 q^2 + 1408 q^3 + 6144 q^4 + \cdots
\een
There will be a unique solution to the spectral flow equation which satisfies these boundary data. To show the uniqueness, we need to show that the system of equations
\bea
h_{-1}(\tau)\theta_{(-1)}(\tau,0)+h_0(\tau)\theta_{(0)}(\tau,0) &=& Z(\tau,0)\cr
h_{-1}(\tau)\theta_{(-1)}(\tau,1/2)+h_0(\tau)\theta_{(0)}(\tau,1/2) &=& Z(\tau,1/2)
\eea
has a unique solution $(h_{-1}(\tau),h_0(\tau))$. To this end, we note that
\bea
\theta_{(-1)}(\tau,1/2) &=& -\theta_{(-1)}(\tau,0)\cr
\theta_{(0)}(\tau,1/2) &=& \theta_{(0)}(\tau,0)
\eea
which can be seen from the following product realizations,
\bea
\theta_{(0)}(\tau,z) &=& \theta_3 (2\tau,2z) =\prod^\infty_{m=1} (1-q^{2m}) (1+y^2 q^{2m-1}) (1+y^{-2} q^{2m-1})\cr
\theta_{(-1)}(\tau,z) &=& \theta_2 (2\tau,2z) =q^{\frac{1}{4}}(y+y^{-1})\prod^\infty_{m=1} (1-q^{2m}) (1+y^2 q^{2m}) (1+y^{-2} q^{2m})
\eea
In principle then, we can proceed and express the solutions in a rather complicated by solving the above equations for $h_{-1}(\tau)$ and $h_{0}(\tau)$. However, the form of this solution will be rather complicated. It turns out that the solution to the spectral flow equation and the above boundary data can be expressed in a simple form as
\ben
Z(\tau,z)&=& \frac{1}{2} \left[
\left(\frac{\theta_3 (\tau,z)}{\theta_3 (\tau,0)}\right)^2 + \left(\frac{\theta_4 (\tau,z)}{\theta_4 (\tau,0)}\right)^2
\right] \cr
&=& 1+ (14+y^2+ y^{-2}-8(y+y^{-1}))q 
\cr
&+& (100+14(y^2+ y^{-2})-64(y+y^{-1}))q^2 + \cdots
\label{egtn}
\een
To show this, it is enough to check that this solutions satisfies the boundary data (which is obvious) and the spectral flow equation. It then follows from uniqueness that this must be identical with the solution that we find by solving for $h_{-1}$ and $h_{0}$ and plugging this back into $Z(\tau,z) = h_{-1}(\tau) \theta_{(-1)}(\tau,z)+h_{0}(\tau) \theta_{(0)}(\tau,z)$. To show that the solution (\ref{egtn}) satisfies the spectral flow equation 
\bea
Z(\tau,z+m\tau+n) &=& e^{-2\pi i \(m^2\tau + 2m z\)} Z(\tau,z)
\eea
we use Eqs (\ref{periodb}), (\ref{perioda}). One may also check explicitly that the first few terms in the series expansions agree for the two different forms of this solution.

One can see that the Witten index $Z(\tau,0)=1$ is invariant under the SL(2,$\mathbb{Z}$) modular transformation as 
required by the RR sector with the fermion boundary condition required by $(-1)^F$ term along the time circle direction.

Including $z$ dependence, one finds that 
\bea
Z(\tau+1,z)&=&Z(\tau,z)\cr
Z(-1/\tau, z/\tau) &=& e^{\pi i   \hat{c}\frac{  z^2}{\tau }}Z'(\tau,z)
\eea 
where 
\be
Z'(\tau,z)= \frac{1}{2} \left[
\left(\frac{\theta_3 (\tau,z)}{\theta_3 (\tau,0)}\right)^2 + \left(\frac{\theta_2 (\tau,z)}{\theta_2 (\tau,0)}\right)^2
\right]
\ee
Hence we do not have a covariance under the full  SL(2,$\mathbb{Z}$) modular transformation. 

But under the modular transformation by an element of $\Gamma_0 (2)$,  the above elliptic genus is covariant as
\be
Z\Big(
\frac{a\tau +b}{c \tau +d}, \frac{z}{c\tau+d}
\Big) =e^{ \pi i  \hat{c}\frac{ c z^2}{c\tau +d}}Z(\tau,z)
\ee
Note that the usual SL(2,$\mathbb{Z})$ is defined by
\be
\tau \rightarrow \frac{a\tau +b}{c\tau +d}
\ee
where $ad-bc=1$ and $a,b,c,d \in \mathbb{Z}$. Then $\Gamma_0(N)$ is a subgroup of SL(2,$\mathbb{Z}$) 
with further conditions
 $a,d=1 \ {\rm mod} \ N$ and $c=0\  {\rm mod} \ N $.

To show this modular property, we note
\bea
&& Z(\tau+1,z)=Z(\tau,z) \cr
&& Z\Big(\frac{-\tau}{2\tau-1}, \frac{z}{2\tau-1}\Big) = e^{\pi i  \hat{c}\frac{ 2 z^2}{2\tau -1}}Z(\tau,z)
\eea 
where the first and the second are respectively  $T$ and $S T^2 S$ which are generators of $\Gamma_0(2)$.

\subsubsection{The refinement}
The (4,4) AH sigma model has a SO(3) global symmetry in addition to the SU(2) R symmetry  
discussed previously. Let us here denote the Cartan generator of the SO(3) isometry by $J_{\rm U(1)}$, which is proportional to the momentum $Q_{relative}$ which is conjugate to the periodic fiber-coordinate of AH. The proportionality constant is yet to be determined. We shall compute the refined index defined by
\be
\h Z (\tau, z,\mu)= {\rm Tr}_{\rm RR} (-1)^{F_L+F_R} x^{J_{\rm U(1)}} y^{J_L} q^{L_0 -\frac{c}{24}} {\bar{q}}^{\bar{L}_0 -\frac{c}{24}} 
\ee 
where $x=e^{2\pi i \mu}$. Obviously,
 \be
\h Z(\tau, z,0)=Z(\tau,z)
\ee
where $Z(\tau,z)$ is the elliptic genus for AH given in (\ref{egtn}).

We note that the action of $J_{U(1)}$ is similar to $J_L$ but it acts both 
fermions and bosons at the same time whereas $J$ acts only on fermions.
Since $J_L$ and $J_{U(1)}$ commute with each other, they are simultaneously diagonalized. Also 
within each BPS multiplet, any positively charged state should be paired with a negatively charged
state with the same magnitude of the charge. 

We may also note that even though $\theta_1(\tau,-z) = -\theta_1(\tau,z)$ is odd, the three remaining theta functions are even, so it does not matter whether we use $\theta_{2,3,4}(\tau,z)$ or $\theta_{2,3,4}(\tau,-z)$ since they are equal. 

With these considerations in mind, there are basically two possibilities of the refined elliptic genus. One is 
\be
\h Z (\tau,z,\mu)= \frac{1}{2} \left[
\frac{\theta_3 (\tau,z+\mu)\theta_3 (\tau,z-\mu)}{\theta^2_3 (\tau,\mu)} + 
\frac{\theta_4 (\tau,z+\mu)\theta_4 (\tau,z-\mu)}{\theta^2_4 (\tau,\mu)}
\right]
\label{reg}
\ee
The other is
\be
\h Z' (\tau,z,\mu)= \frac{1}{2} \left[
\left(\frac{\theta_3 (\tau,z+\mu)}{\theta_3 (\tau,\mu)}\right)^2 + 
\left(\frac{\theta_4 (\tau,z+\mu)}{\theta_4 (\tau,\mu)}\right)^2
\right]
\ee
Below we shall show that the former is the correct one. 
Let us first note that, for either possibilities,
\be 
\h Z (\tau,0,\mu)=1
\label{d1}
\ee
and 
\be
\h Z (\tau,1/2,\mu)= \frac{1}{2} \left[
\left(\frac{\theta_4 (\tau,\mu)}{\theta_3 (\tau,\mu)}\right)^2 + 
\left(\frac{\theta_3 (\tau,\mu)}{\theta_4 (\tau,\mu)}\right)^2
\right]
\label{d2}
\ee
Since $J_{U(1)}$  is integral or half-integral quantized for any boundary conditions, 
$J_{U(1)}$ is independent of the spectral flow.
Therefore,
\be
\h Z (\tau, z+m \tau+n,\mu)= (-1)^{\hat{c}(m+n)} e^{-\pi i \hat{c} ( m^2 \tau + 2 m z)}\, \h Z (\tau, z,\mu)
\ee
Then with $\hat{c}=2$, the most general solution reads
\be
\h Z (\tau,z,\mu)= H_{-1}(\tau,\mu) \theta_{(-1)}(\tau,z)+H_{0}(\tau,\mu) \theta_{(0)}(\tau,z)
\ee
Using the data (\ref{d1}) and (\ref{d2}), $H_{-1}$ and $H_{0}$ can be determined as
\bea
H_0 (\tau,\mu) &= &\frac{\h Z (\tau,0,\mu)+ \h Z (\tau,1/2,\mu)}{2 \theta_{(0)}(\tau,0)}  \cr
H_{-1}(\tau,\mu) &=& \frac{\h Z(\tau,0,\mu)- \h Z(\tau,1/2,\mu)}{2 \theta_{(-1)}(\tau,0)} 
\eea
We have then confirmed that the first few terms in a series expansion agree with (\ref{reg}). We may also verify that only this solution solves the spectral flow equation since this is the only option where the $\mu$ dependence among the phase factors
\bea
\theta_3(\tau,z \pm \mu + \tau) &=& e^{-\pi i \tau - 2\pi i (z\pm \mu)} \theta_3(\tau,z\pm \mu)\cr
\theta_4(\tau,z \pm \mu + \tau) &=& e^{-\pi i \tau - 2\pi i (z\pm \mu)} \theta_4(\tau,z\pm \mu)
\eea
is canceled. 

The refined elliptic genus has the transformation rule
\bea
\h Z\(\frac{a\tau +b}{c \tau +d}, \frac{z}{c\tau+d}, \frac{\mu}{c\tau+d}\) &=& e^{\pi i  \hat{c}\frac{ c z^2}{c\tau +d}} \h Z(\tau,z,\mu)
\eea
which defines a weak Jacobi form of weight zero and index one under $\Gamma_0(2)$. 

Around $x=0$ we have a Laurent expansion on the form
\bea
\h Z(\tau,z,\mu) &=& Z_0^{\B}(\tau,z) + \sum_{n=1}^{\infty} \(x^{2n}+x^{-2n}\) Z_{2n}(\tau,z)
\eea
where the first few Fourier coefficients have the $q$-expansions 
\bea
Z_0^{\B}(\tau,z) &=& 1+\frac{(1-y)^4}{y^2}q+\frac{6(1-y)^4}{y^2}q^2-\frac{4(1-y)^4(1-(5-y)y)}{y^3}q^3 + \Ordo(q^4)\cr
Z_2(\tau,z) &=&-\frac{2(1-y)^2}{y}q+\frac{4(1-y)^4}{y^2}q^2-\frac{2(1-y)^2(1-(1-y)y(10-(9-y)y))}{y^3}q^3 + \Ordo(q^4)\cr
Z_4(\tau,z) &=& -\frac{4(1-y)^2}{y}q^2+\frac{8(1-y)^4}{y^2}q^3 + \Ordo(q^4)\cr
Z_6(\tau,z) &=&-\frac{6(1-y)^2}{y}q^3 + \Ordo(q^4)\cr
Z_8(\tau,z) &=& \Ordo(q^4)
\eea

It now seems to us that we shall identify $J_{\rm U(1)}=2Q_{relative}$ and then we read off the bound 
\ben
|Q_{relative}|\leq k\label{bindning}
\een
on the possible relative charges on these left-moving BPS states for a given momentum $k/R_5$. With this choice of normalization, all left-moving BPS states carry integer values on $Q_{relative}$. 

We see that the expansion of the zero-charge component, $Z^{\B}_0$, is in agreement with the single particle index $Z^{\A}_0$ in theory $\A$ which we presented in eq (\ref{Z0expansion}) and which was extracted from the U(3) dyonic instanton index. 

The sum of all nonzero (positive as well as negative) charge contributions (at $x=1$) is 
\ben
Z^{\B}_{charged}(\tau,z) &=& 0 + \(8-4(y+y^{-1})\)q + \(64-40(y+y^{-1})+8(y^2+y^{-2})\)q^2\cr
&& + \Ordo(q^3)\label{charged}
\een
Multiplying this by the center of mass contribution (\ref{su2}) we find an agreement with the expansion (\ref{orderq2}). 

At $y=-1$ we have
\bea
Z^{\B}_0(\tau,1/2) &=& 1+16q+96q^2+448q^3 + \cdots\cr
Z^{\B}_{charged}(\tau,1/2) &=& 0+16q+160q^2+950q^3+\cdots
\eea
and these expansions are consistent with $Z_0^{\A}$ presented in (\ref{high1}) and with $z'_{N=2,n=2}/z'_{N=2,n=1}$ from eq (\ref{high2}) respectively.

\section{Two identical SU(2) monopole strings}\label{section7}
The moduli space of two identical massive fundamental monopole strings in U(2) 5d MSYM is 
\bea
M_2 &=& \mb{R}^3 \times \frac{S^1 \times M_{AH}}{\mb{Z}_2}
\eea
where $M_{AH}$ denotes the Atiyah-Hitchin space. We now note the $\mb{Z}_2$ projection on the total moduli space $M_2$. This ${\mb{Z}}_2$ acts on the $2\pi$ ranged $S^1$ coordinate $\chi$ as 
\bea
\chi &\rightarrow \chi + \pi
\eea
The momentum conjugate to this coordinate is the total electric charge. The $\mb{Z}_2$ acts simultaneously on the AH space \cite{Sen:1994yi} (Eq.~(15)) as a translation of a $2\pi$ ranged relative angle in AH space,
\bea
\psi \rightarrow \psi + \pi
\eea
Also note that the latter transformation alone is the discrete isometry of AH space, for which one may refine the elliptic genus. We shall 
denote this symmetry of AH space by $\mathbb{Z}^{AH}_2$.

The story for the zero mode part is well known  \cite{Sen:1994yi}.
Since the monopole charge quantum number is $2$ as we consider two monopole strings, the electric
charge quantum number for any threshold bound state is an odd integer $2n+1$ which is relative 
prime to the magnetic charge. The wave function for a dyonic string with electric charge quantum number $2n+1$ 
on the center of mass part of the moduli space is
\bea
\psi_{com} &=& e^{i(2n+1) \chi}
\eea
and it picks up the factor of $-1$ under the above $\mb{Z}_2$ transformation. The total wave function shall be invariant under $\mb{Z}_2$ so the relative part of the wave function $\psi_{rel}$ which lives on the AH space, must likewise pick up a factor of $-1$ under $\mb{Z}_2$. In \cite{Sen:1994yi} it was shown that there exist such a wave function. It was also shown that there does not exist a zero-mode wave function on AH which is even under $\mb{Z}^{AH}_2$.

There will be a corresponding separation of the elliptic genus on AH into odd and even parts under the $\mb{Z}^{AH}_2$, 
\bea
Z_{AH} &=& Z_{AH,even} + Z_{AH,odd}
\eea
where $Z_{AH,odd}$ corresponds to insertion of the projector $\frac{1}{2}(1-g)$ inside the trace that defines the elliptic genus, where $g$ implements the $\mb{Z}^{AH}_2$ action on the AH space (and likewise, $Z_{AH,even}$ corresponds to inserting the projector $\frac{1}{2}(1-g)$ inside the trace). It is thus the odd part $Z_{AH,odd}$ which contains 
a contribution from a threshold bound state at zero momentum $q^{k = 0}$. In other words, this part of the elliptic genus starts as 
\bea
Z_{AH,odd} &=& 1 + \Ordo(q)
\eea
and corresponds to electrically charged monopole strings. Furthermore, there is no threshold bound state which is 
even under $\mb{Z}^{AH}_2$ so we shall have 
\bea
Z_{AH,even} &=& 0 + \Ordo(q)
\eea
In this paper we are concerned with monopole strings that carry no electric charge. Hence what we are after is $Z_{AH,even}$.

From the dyonic instanton index, we could extract from the single-particle index, the contribution that comes from two identical SU(2) monopole strings as
\bea
z'_{N=2,n=2}(q,y) &=& \(8-4(y+y^{-1})\) q + \(112 -72(y+y^{-1})+16(y^2+y^{-2})\) q^2 + \Ordo(q^3)
\eea
We now recall that this expression also agrees with $z_{N=2,n=1}(q,y) Z^{\B}_{charged}(q,y)$ that we obtained from the elliptic genus on AN. (The factor $z_{N=2,n=1}$ corresponds to the center of mass contribution coming from $\mb{R}^3 \times S^1$ part of the monopole moduli space.). 
We see that $Z_{charged}(AN)$ corresponds 
to $Z_{even}(AH)$ which in turn corresponds to the elliptic genus of two identical, and electrically neutral SU(2) monopole strings.

\section{Higher rank U(N) gauge groups}\label{zerocharge}
We may deform the $(4,4)$ sigma model by a potential without breaking any supersymmetry. However such a potential will modify the superalgebra by extra central charge terms\footnote{\label{tio}For instance, if we (for the sake of simplicity of our illustration) select the $(1,1)$ 
supercharge sector and add a central charge $Z$ there, then we will get the superalgebra
\bea
\{\mb{Q},\mb{Q}\} = \(\begin{array}{cc}
\{Q_+,Q_+\} & \{Q_+,Q_-\}\\
\{Q_-,Q_+\} & \{Q_-,Q_-\}
\end{array}\) &=& 
2\(\begin{array}{cc}
H-P & -Z\\
-Z & H+P
\end{array}\)
\eea
where the $(1,1)$ supercharges $Q_{\pm}$ are real, and hence the supercharge matrix $\mb{Q}$ is hermitian, enabling us to write the left-hand side as $\{\mb{Q},\mb{Q}^{\dag}\}$ which is manifestly non-negative. To find the BPS energy $H$, we then need to find the eigenvalues of the matrix
\bea
\(\begin{array}{cc}
-P & -Z\\
-Z & +P
\end{array}\)
\eea
Its eigenvalues are $\pm \sqrt{P^2 + Z^2}$ and therefore the BPS bound is given by 
\bea
H &=&  \sqrt{P^2 + Z^2}
\eea}.
These central charge terms will correspond to U(1) charges which are Lie derivatives along U(1) isometries on the moduli space. If a state has non-vanishing central charge it will not give a net contribution to the elliptic genus. Only right-movers satisfying $H=P$ contribute. Thus if we deform the sigma model so that some states have non-vanishing central charge, this deformation will deform our elliptic genus. Our first priority would be to find a  deformation which does not deform our elliptic genus. This turned out to be a difficult task. We were unable to find any such deformation. However if we are interested in the zero U(1) charge sector of the elliptic genus only, then we may indeed tolerate that the elliptic genus can be deformed, since the zero charge sector will be kept intact by such a deformation. This is because the BPS equation will remain $H=P$ on those zero charged BPS states. 

The bosonic part of the potential produced by our deformation \cite{Bak:1999da,Bak:1999ip}
is given by 
\bea
V &=& g_{rs} G^r G^s
\eea
where 
\bea
K &=& G^s \partial_s 
\eea
is a tri-holomorphic Kiling vector field on the monopole moduli space. Let us begin with $U(3)$ and consider only Taub-NUT part of the moduli space with metric
\bea
ds^2 &=& \(1+\frac{1}{r}\) \(dr^2 + r^2 \(\sigma_1^2 + \sigma_2^2\)\) + \frac{1}{1+\frac{1}{r}} \sigma_3^2
\eea
Here we have a tri-holomorphic vector field 
\bea
K &=& a \partial_{\psi}
\eea
where $\sigma_3 = d\psi + \cos \theta d\phi$ and $\psi$ is $4\pi$ periodic. The potential is 
\bea
V &=& a^2 g_{\psi\psi} = a^2 \frac{r}{r+1}
\eea
Near the minimum $r=0$ the Taub-NUT space is flat $\mb{R}^4$ with the flat metric
\bea
ds^2 &=& \frac{1}{r} dr^2 + r \(\sigma_1^2 + \sigma_2^2 + \sigma_3^2\)
\eea
We may define $r=\rho^2/4$ for which this metric becomes
\bea
ds^2 &=&  d\rho^2 +\frac{\rho^2}{4}  \(\sigma_1^2 + \sigma_2^2 + \sigma_3^2\)
\eea
The potential is that of the harmonic oscillator near $\rho = 0$,
\bea
V &=& a^2 \rho^2
\eea
By taking $a$ large the wave function localizes near $\rho=0$ where Taub-NUT space can be replaced with $\mb{R}^4$. 

Once we have localized to the origin of TN, we may turn off the potential again since the zero charge sector of the elliptic genus does not depend on $a$, no matter the potential is turned on for sigma model on $\mb{R}^4$ or TN. We thus expect that the zero charge sector of the elliptic genus for TN and for $\mb{R}^4$ coincide. This explains the result that we presented in eq (\ref{eg10}).

For higher rank U(N) gauge group we again project onto the sector where all the relative charges are zero. The full elliptic genus factorizes as
\bea
Z &=& Z_{com} Z_{rel}
\eea
This full elliptic genus is not known for $N=4,5,\cdots$. But we may again introduce a potential picking up contributions for each relative electric charge in the relative part of the moduli space. Taking the vev's to be large, the wave function localizes to the origin of the moduli space where it is locally on the form $\mb{R}^{4(N-2)}$ which we can think of as $N-2$ copies of harmonic oscillators on $\mb{R}^4$. From each $\mb{R}^4$ we pick up the contribution $Z^{\B}_0$. Hence the full contribution from the zero charge sector to the elliptic genus becomes
\bea
Z &=& Z_{com} (Z^{\B}_0)^{N-2}
\eea
Indeed this is the structure we find from the dyonic-instanton index computation given in Eq.~(\ref{Z0}).

\section{The duality between theory $\A$ and theory $\B$}
We have argued that duality explains why we may identify the elliptic genus of the monopole string in the zero charge sector with the dyonic instanton index. We would now like to explain this duality in more detail. First this duality can be understood quite intuitively from the M5 brane viewpoint as two different dimenional reductions. Unfortunately we do not know a direct formulation of the M5 brane theory and therefore this argument cannot be made very precise. Nevertheless, this provides a quite simple geometrical picture of the duality. We will therefore begin by describing the duality as dimensional reductions of M5. We will then also describe the duality using string theory S- and T-dualities.

\subsection{Duality viewed from M5 brane}
Let us first consider the U(2) case ($N=2$) since this involves only the overall part of the moduli space of the corresponding monopole string. The overall part of the dynamics is described by the 2d sigma model with a flat target space $\R^3 \times S^1_e $. Note that the $S_e^1$ here is the gauge circle direction and there is a discrete target-space momentum $Q_{overall}$ along the $S_e^1$ direction. From the viewpoint of the 5d MSYM theory, this target-space momentum $Q_{overall}$  implies the presence of an electric field transverse to the monopole string and along the 123 directions, which is generated by the electric charge distributed along the monopole string location\footnote{The F1(06) is dissolved into the D2(056) which ends on D4(01235) and the electric charge is therefore homogeneously distributed along the monopole string.}. This is a dyonic string where $Q_{overall}$ represents the total charge of  F1 (06) stretched between D4 branes. In the dual side, this F1 (06) corresponds to the D2 (046) connecting D4 branes. 

The target-space momentum $Q_{overall}$ is carried only by the zero mode part of the 2d (4,4) sigma model, which is clear from the definition
\be
Q_{overall}= \frac{1}{2\pi}\int_0^{2\pi} \, d\sigma \dot{X}_e (t, \sigma)
\ee 
where $X_e$ is the target-space coordinate of the $S_e^1$ direction.
Namely the target-space momentum $Q_{overall}$ does not receive any contribution from the oscillator part of the sigma model. 

Let us now consider the M-theory brane configuration
\bea
\begin{array}{lll}
N=2 & M5 & (012345)\\
n & M2 & (056)\\
Q_{overall} & M2 & (046)\\
k & W & (05)
\end{array} 
\eea
where $4$ and $5$-directions are along a two-torus. We assume the M5 branes are separated by a vev $v$ in the 6th direction.
If we dimensionally reduce along $5$-th direction, we get theory $\A$ with $k$ dyonic instantons carrying electric charge $n$ and supplemented with $Q_{overall}$ D2's
\bea
\begin{array}{lll}
N=2 & D4 & (01234)\\
n & F1 & (06)\\
Q_{overall} & D2 & (046)\\
k & D0 & (0)
\end{array} 
\eea
If we instead dimensionally reduce along the $4$-th direction, we get theory $\B$ with $n$ monopole strings carrying electric charge $Q_{overall}$ and wave with momentum $k/R_5$,
\bea
\begin{array}{lll}
N=2 & D4 & (01235)\\
n & D2 & (056)\\
Q_{overall} & F1 & (06)\\
k & W & (05)
\end{array} 
\eea
The corresponding M5 brane superalgebra reads
\bea
\{Q,Q^{\dag}\} &=&M + \Gamma^{05} \frac{k}{R_5} + \Gamma^{056} n M_{056} + \Gamma^{046} Q_{overall} M_{046}
\eea
Here $M_{056}$ denotes the mass of a single M2 brane (056) and $M_{046}$ denotes the mass of a single M2 (046). Since $\Gamma^{046}$ and $\Gamma^{056}$ anti-commute, and they both commute with $\Gamma^{05}$, we get the BPS mass \cite{Townsend:1997wg} 
\bea
M &=& \frac{k}{R_5} + \sqrt{n^2 M_{056}^2+ Q_{overall}^2 M_{046}^2}
\eea
If $Q_{overall}=0$ this is the usual mass formula of a 1/4-BPS dyonic instanton (M2-W threshold bound state),
\ben
M &=& \frac{k}{R_5} + n M_{056}\label{bound2}
\een
where $n$ is the electric charge and $k$ the instanton number. It is also the energy of a wave with momentum $P = \frac{k}{R_5}$ along the monopole strings in theory $\B$. 

For the 2d sigma model effective field theory, to get the energy of the wave on the monopole string, we need to subtract the mass $M_{monopole} = nM_{056}$ of the static monopole strings. We are then left with the mass of the wave along the monopole strings,
\bea
H &=& M - M_{monopole} 
\eea
which will correspond to the Hamiltonian of the 2d sigma model. If we expand this out, we get
\bea
H &=& \frac{k}{R_5} + \frac{Q_{overall}^2 M_{046}^2}{2 n M_{056}}
\eea
in the non-relativistic limit (which is the limit in which the 2d sigma model is defined). This mass exceeds the momentum $P = \frac{k}{R_5}$ if $Q_{overall}$ is non-zero. States which contribute to our elliptic genus have $H=P$ and therefore states with $Q_{overall}\neq 0$ do not contribute. We note that this is true for both $k=0$ and $k>0$.

For U(N) gauge group when $N >2$ we have also a relative part of the moduli space. Let us consider $N=3$ to be specific, and where the relative part of the moduli space is the TN space \cite{Lee:1996if,Gauntlett:1996cw}. Let us assume generic electric charges $q_{12}$ and $q_{23}$. Here $q_{12}$ counts the number of oriented F1 stretched between D4$_1$ and D4$_2$ which can be positive or negative integer number or zero.  Likewise for $q_{23}$. The momentum along the overall gauge direction (which generically is no longer a circle) is given by  
\ben
Q_{overall} &=& \frac{v_{12} q_{12} + v_{23} q_{23}}{v_{12} + v_{23}}\label{Qoverall}
\een
The quantization we see in $Q_{overall}$ corresponds to the $\mb{Z}$-identification on the monopole moduli space (\ref{Zid}). The momentum along the TN fiber-circle is half-integer quantized due to the $4\pi$-identification of the fiber-circle and is given by
\ben
Q_{relative} &=& \frac{q_{12} - q_{23}}{2}\label{Qrelative}
\een
The mass of the F1 strings is given by
\bea
M_{046} &=& \frac{1}{2\pi} \(|v_{12}q_{12}| + |v_{23}q_{23}|\)
\eea
Just as for the U(2) case, here again states with nonvanishing $Q_{overall}$ can not be left-moving, and therefore they do not contribute to the elliptic genus. When $v_{12} = v_{23}$ and $Q_{overall} = 0$, we find that $Q_{relative} = q_{12}$ is integer quantized. When $v_{12} \neq v_{23}$ it is generically  not possible to put $Q_{overall} = 0$ unless also $Q_{relative} = 0$ which means that generically only states with $Q_{overall} = Q_{relative} = 0$ contribute to the elliptic genus. However when $v_{12} = v_{23}$ states with nonvanishing $Q_{relative}$ will contribute. Now when we computed the elliptic genus on TN we did not take into account the $\mb{Z}$-identification on the full monopole moduli space. We then got an elliptic genus on TN where some states appear to carry nonvanishing $Q_{relative}$. To obtain the elliptic genus of the full monopole moduli space, we have to project out states which are killed by the $\mb{Z}$-identification. For generic vev's they are all killed, although for exceptional cases, some such states may survive, and if $v_{12}=v_{23}$ they will all survive.

Momentum $Q_{relative}$ along the TN space leads to an increased kinetic energy $H$ of the 2d sigma model Hamiltonian. When $k=0$ the momentum $Q_{relative}$ does not induce an increased longitudinal momentum $P$ and therefore these states are not left-moving.  Therefore states with $M_{046}>0$ and $k=0$ do not contribute to the elliptic genus. 

On the other hand, when $k>0$ the TN space momentum $Q_{relative}$ does induce a corresponding increased momentum $P$. So the bound of the elliptic genus $H=P$ can be still satisfied. We can perform an explicit computation for a sigma model on $\mb{R}^4 = \mb{C}^2$ which is how TN space looks near its origin. There momentum in TN corresponds to common phase rotation in the both factors $\mb{C}^2 \times \mb{C}^2$. We may thus consider sigma model with target $\mb{C}$ since the two factors are decoupled. Here it is straightforward to verify that $H=P$ can hold for states with a nonvanishing U(1) charge (where the U(1) acts on the phase of $\mb{C}$), as can also be seen explicitly by expanding out the final result that we presented in Eq (\ref{u1}).

More generally, for $N>2$, we have in place of TN the general LWY moduli space metric, which has U(1)$^{N-2}$ isometries. (It also has an SU(2) isometry, just like the TN space has.) The corresponding momenta along these U(1) directions are relative electric charges, and, for $k>0$, we may excite these relative U(1) charges while preserving the equation $H=P$. So such charged states do contribute to our elliptic genus.

\subsection{Duality viewed from D3 brane}\label{TST}
We have no direct access to the M5 brane theory, and therefore the argument we presented in the previous section cannot be made very precise. Instead we can use TST-duality. We use T-duality along the 4th direction and map theory $\A$ on the D4 branes to D3 branes in type IIB string theory. Here we can use S-duality to get another D3 brane theory. Finally we may T-dualize back to get the other D4 brane theory $\B$. 

In theory $\A$ we have dyonic-instanton BPS states which can be realized in IIA string theory as the brane configuration of $N$ D4 (01234), $k$ D0 (0) and F1 (06). The projection operators of surviving supersymmetries are
\be
\Pi_{\rm D4}=\frac{1+\Gamma_{012345}}{2}, \ \Pi_{\rm D0}=\frac{1+\Gamma_{05}}{2}, \ \Pi_{\rm F1}=
\frac{1+\Gamma_{065}}{2}
\label{projectionins}
\ee
where $\Gamma_5$ is the 10d chirality matrix 
\bea
\Gamma_5 &=& \Gamma_{012346789(10)}
\eea
but we may also think on this in the M-theory context as the gamma matrix of a 5th direction which will be the M-theory circle. Since all three projections (\ref{projectionins}) mutually commute, we have 4 real remaining supersymmetries after the projection down to the 1/4-BPS states\footnote{With D4, we have 16 supersymmetries and 
one 1/4 refers to the fraction of preserved supersymmetries starting from the 16.}. We assume that the 4th direction is circle compactified. Still the instanton configuration may carry angular momentum in the 123 plane. Including the F1, we have dyonic instantons, which again may  carry angular momentum in the 123 plane. When the 4th direction is circle compactified, we shall also include D2 (046). 

Let us now apply the TST duality. First we apply a T-duality along the 4th circle direction. Then D4 turns into D3 (0123), D0 becomes D1 (04) while 
F1 (06) remains. These are the 1/4-BPS dyonic caloron configurations.  They carry an angular momentum along 
the 123 plane.   If there are D2's in addition, they become D1's (06) under the T-duality. The 5d MSYM coupling constant $g_{YM}^2 = 4\pi^2 R_5$ is under T-duality mapped into the 4d MSYM coupling constant 
\bea
G_{YM}^2 &=& \frac{2\pi R_5}{R_4}
\eea

We now apply the S-duality. The D3 (0123) remains, but as we will explain more fully in a moment, since S-duality permutes the 4th and 5th directions, it takes D1 (04) into F1 (05), while  and F1 (06) turns into D1 (06). This configuration corresponds to 1/4-BPS dyons. Again these dyons may involve the angular momentum along the 123 plane, and, if there are D1's (06) in addition, they become F1's (06). S-duality maps the U(N) gauge group into its Langlands dual, which is again U(N). It maps the coupling constant into its inverse
\bea
G_{YM}^2 &\rightarrow & G_{YM}'^2 = \frac{4\pi^2}{G_{YM}^2}
\eea
or if we define $\tau = \frac{4\pi i}{G_{YM}^2}$, then S-duality maps $\tau \rightarrow -1/\tau$. Thus we obtain after S-duality the coupling constant
\bea
G_{YM}'^2 &=& \frac{2\pi R_4}{R_5}
\eea
In this paper we take the passive viewpoint that S-duality acts on $\tau$ while we keep the torus and $R_4$ and $R_5$ fixed. But we could also have taken the active viewpoint and let the transformation instead act on the torus coordinates $x^4$ and $x^5$ by a large diffeomorphism $x^4 \rightarrow x^5$ and $x^5 \rightarrow -x^4$ which permutes $R_4$ and $R_5$. Both viewpoints lead to the same transformation of the 4D SYM coupling.

Finally we apply again T-duality along the 5th circle direction. Then D3 becomes D4 (01235), D1 becomes D2 (056), while, for F1 (05), 
the winding and 
the momentum along the 5th circle direction will be exchanged.  Namely the winding (corresponding to the instanton number 
$k$  in our original configuration) 
becomes the momentum $P(=k)$ along 
the 5th circle direction.
The corresponding projections read 
\be
\Pi_{\rm D4}=\frac{1+\Gamma_{012345}}{2}, \ \Pi_{P}=\frac{1+\Gamma_{05}}{2}, \ \Pi_{\rm D2}=
\frac{1+\Gamma_{056}}{2}
\ee
which commute one another leading to 1/4-BPS configurations.
 In case there are F1's (06) in addition, they remain to be F1's (06) under the second T-duality. The coupling constant of this 5d MSYM theory (theory $\B$) becomes $g_{YM}'^2 = 4\pi^2 R_4$. This was to be expected from the M5 brane viewpoint where we dimensionally reduce along the 4th direction to get theory $\B$. The gauge group remains U(N).

The decompactification limit $R_4 \rightarrow \infty$ of theory $\A$ corresponds to the strong coupling limit of theory $\B$. But to compute the elliptic genus we need to take the weak coupling of theory $\B$. However, the elliptic genus depends on $q = \exp 2\pi i \tau$ where $\tau = i\beta/R_5$. Hence $q$ is independent of the SYM coupling $R_4$ in theory $\B$. Furthermore, the elliptic genus is a topological invariant, so it will not change at all if we continuously change from strong coupling ($R_4 \gg R_5$) to weak coupling ($R_4 \ll R_5$). We can thus compute the elliptic genus at weak coupling, and then compare with the dyonic instanton index by a trivial extrapolation of our elliptic genus to strong coupling.

In the decompactification limit of theory $\A$ the number of D2 (046) correspond to various superselection sectors. In \cite{Kim:2011mv} the sector with no D2 (046) was considered in the decompactification limit. 

In the dual theory $\B$ this corresponds to the projection of the elliptic genus down to the zero charge sectors. For the overall part we do not have the concept of relative charge. Thus the projection on the elliptic genus becomes on the form 
\be
Z= \(Z_{\rm com}\) \times \(Z_{\rm rel}|_{Q_{relative}=0}\)
\ee
This is then the quantity which we have matched with the result in \cite{Kim:2011mv} of theory $\A$.

\section{Discussion}
We have seen that duality between theory $\A$ and theory $\B$ can be argued to be a TST-duality, and one may think that we have added nothing new to already well-known dualities in string theory. However, we used T-duality to relate D4 to D3 and this does not correspond to dimensional reduction of 5d MSYM to 4d MSYM, but rather to dimensional reduction with all the KK modes kept \cite{Taylor:1996ik}. Therefore our S-duality lives in 5d rather than in 4d since T-duality is mainly a reformulation of the 5d theory. The 5d S-duality from the field theory point of view, has only quite recently been studied \cite{Tachikawa:2011ch}. Our result provides one further evidence of 5d S-duality. It may be seen as an extension to include the KK modes, of the corresponding S-duality checks in 4d MSYM for 1/2-BPS dyon states \cite{Lee:1996if,Gauntlett:1996cw} for U(3) case, and also of \cite{Sen:1994yi} for the U(2) case with two identical monopole strings.\footnote{To be more precise, we checked the S-duality on the zero charge sector $Q_{overall} = Q_{relative} = 0$ only. It will be very interesting to include non-vanishing charged states in our S-duality check. The S-duality of charged states can be tested once we can compute the index of a periodic dyonic-instanton bound to monopole strings.} One may try to extend our S-duality check by also including the 1/4-BPS dyon states \cite{Bak:1999ip} which also will have corresponding dyonic string up-lifts. 

In this paper we did not construct any states explicitly. Their quantum numbers and degeneracy are encoded in our elliptic genera and index on the dual side. But an explicit construction of states would be a nice confirmation of our result. For low values of instanton number $k$ one may hope to be able to explicitly construct such periodic dyonic-instanton-monopole-string bound states.

\section*{Acknowledgement}
We would like to thank Stefan Hohenegger, Seok Kim, Kimyeong Lee, Filippo Passerini and Soo-Jong Rey for discussions, and for correspondence with Jan Troost which eventually made us aware of our mistake in the first version of this paper. We also thank Nigel Hitchin for providing us with a reference for the Euler characteristic and the signature of AH space. This work was
supported in part by  2013 sabbatical year research grant by University of Seoul.

\appendix
\section{Theta functions}\label{theta}
In this appendix, we present the theta functions and their transformation properties under SL(2,$\mathbb{Z}$). Their basic definitions are as follows:
\bea
\theta_1(\tau,z)&=& i \sum^\infty_{n=-\infty} (-1)^n q^{\frac{1}{2}{(n-1/2)^2}} y^{n-1/2}\cr
&=&2 q^{\frac{1}{8}} \sin \pi z \prod^\infty_{n=1} (1-q^n) (1-y q^n)(1-y^{-1}q^n) \cr
\theta_2(\tau,z)&=& \sum^\infty_{n=-\infty}  q^{\frac{1}{2}{(n-1/2)^2}} y^{n-1/2}\cr
&=&2 q^{\frac{1}{8}} \cos \pi z \prod^\infty_{n=1} (1-q^n) (1+y q^n)(1+y^{-1}q^n)
\eea
and 
\bea
\theta_3(\tau,z)&=& \sum^\infty_{n=-\infty}  q^{\frac{1}{2}{n^2}} y^{n}\cr
&=& \prod^\infty_{n=1} (1-q^n) (1+y q^{n-1/2})(1+y^{-1}q^{n-1/2}) \cr
\theta_4(\tau,z)&=& \sum^\infty_{n=-\infty} (-1)^n q^{\frac{1}{2}{n^2}} y^{n}\cr
&=& \prod^\infty_{n=1} (1-q^n) (1-y q^{n-1/2})(1-y^{-1}q^{n-1/2}) 
\eea
They have the periodicity properties
\ben
&& \theta_1(\tau,z+\tau) = - e^{-\pi i \tau -2\pi i z} \theta_1(\tau,z)\cr
&& \theta_2(\tau,z+\tau) =\ \ e^{-\pi i \tau -2\pi i z} \theta_2(\tau,z)\cr
&& \theta_3(\tau,z+\tau) =  \ \ e^{-\pi i \tau -2\pi i z} \theta_3(\tau,z)\cr
&& \theta_4(\tau,z+\tau) =- e^{-\pi i \tau -2\pi i z} \theta_4(\tau,z)\label{periodb}
\een
and
\ben
&& \theta_1(\tau,z+1) =- \theta_1(\tau,z)\cr
&& \theta_2(\tau,z+1) =- \theta_2(\tau,z)\cr
&& \theta_3(\tau,z+1) = \ \ \theta_3(\tau,z)\cr
&& \theta_4(\tau,z+1) = \ \ \theta_4(\tau,z)\label{perioda}
\een
Their modular transformation are
\bea
&& \theta_1(\tau+1,z) =e^{\frac{\pi i}{4}}\, \theta_1(\tau,z)\cr
&& \theta_2(\tau+1,z) =e^{\frac{\pi i}{4}}\, \theta_2(\tau,z)\cr
&& \theta_3(\tau+1,z) = \ \ \theta_4(\tau,z)\cr
&& \theta_4(\tau+1,z) = \ \ \theta_3(\tau,z)
\eea
and
\bea
&& \theta_1(-1/\tau,z/\tau) =-i (-i\tau)^{1/2} e^{\pi i z^2/\tau}\, \theta_1(\tau,z)\cr
&& \theta_2(-1/\tau,z/\tau) = \ \ (-i\tau)^{1/2} e^{\pi i z^2/\tau}\ \theta_4(\tau,z)\cr
&& \theta_3(-1/\tau,z/\tau) = \ \ (-i\tau)^{1/2} e^{\pi i z^2/\tau}\ \theta_3(\tau,z)\cr
&& \theta_4(-1/\tau,z/\tau) = \ \ (-i\tau)^{1/2} e^{\pi i z^2/\tau}\ \theta_2(\tau,z)
\eea

\section{Derivation of $Z(\tau,0)$}
Let us first compute $Z(\tau,0)$. We follow the computation in \cite{AlvarezGaume:1986nm}. The elliptic genus at this point is nothing but the Witten index
\be
Z(\tau, 0)= {\rm Tr}_{\rm RR} (-1)^{F_L+F_R} q^{L_0 -\frac{c}{24}} {\bar{q}}^{\bar{L}_0 -\frac{c}{24}} 
\ee 
which is independent of $\beta$ due to its topological nature. ($
\beta$ was introduced in section \ref{beta}.) In the $\beta \rightarrow 0$ limit,
the functional integral splits into an integral over the zero modes and an integral over non constant configurations. 
The latter can be evaluated in the perturbation theory in $\beta$, and its leading term is given by the
ratio of the fermion and bosonic determinants, which comes from the Gaussian approximation of the action of the
non constant modes. Due to the supersymmetry, the leading zero point energy contributions of non-constant boson and fermions
are canceling with each other, leading to
\be
Z_{\rm non}=1+ O(\beta) 
\ee
Thus in the $\beta \rightarrow 0$ limit,  the index has the path integral representation
\be
Z(\tau,0) =\frac{1}{(2\pi)^{\frac{d}{2}}} \int d^d x \sqrt{g} \int \prod^d_{m=1} d\psi^m_{0+} d\psi^m_{0-}
\exp {-\frac{1}{12} R_{ijkl} {\bar{\psi}}^i_{0}     {{\psi}}^k_{0}           {\bar{\psi}}^j_{0}  {{\psi}}^l_{0}}     
\ee
This can be written as
\be
Z(\tau,0) =\frac{1}{(2\pi)^{\frac{d}{2}}} \int d^d x \sqrt{g} \int \prod^d_{m=1} d\chi^{m*}_{0} d\chi^m_{0}
\exp {-\frac{1}{4} R_{ijkl} {{\chi}}^{i*}_0     {{\chi}}^{j*}_0           {{\chi}}^k_0  {{\chi}}^l_0}     
\ee
where we introduced
\be
\chi^m_0 =\frac{1}{\sqrt{2}} \big( 
\psi^m_{0+}+i  \psi^m_{0-}
\big)
\ee
This then becomes 
\be
Z(\tau,0)= \chi(M_d)
\ee
where $\chi$ denotes the Euler  characteristic of the manifold $M_d$.
For case of $d=4$, explicitly
\be
Z(\tau,0)= \frac{1}{32\pi^2} \int_{\rm TN} \epsilon_{abcd} R^{ab} R^{cd} =1
\ee

\section{Derivation of $Z(\tau,1/2)$}
For this computation we use path-integral and we closely follow \cite{Alvarez:1987de}, with emphasis on the regularization problem, which can be avoided at $z=1/2$.
 
For $z=1/2$, we do not have the fermion zero mode for the right moving sector. Hence the expansion 
of the Lagrangian reads
\bea
\bar{L}=\tau_2 \left[
\xi^i \Big( 2\partial_- \partial_+\delta_{ij} -i{R^+_{ij}}\partial_-\Big) \xi^j 
+
 i\zeta_-^i \Big( \partial_+\delta_{ij} -\frac{iR^+_{ij}}{2}\Big) \zeta_-^j
 -i\zeta_+^i \partial_-  \zeta_+^i
\right]
\eea
to the leading order contribution  of $\tau_2$.
The  partition function becomes
\bea
Z(\tau,1/2)= N_d \int d^d x \prod^d_{m=1} d\psi^m_{0+} \ \frac{{\rm det}^{1/2}\Big(
i\tau_2 \big(
\partial_+ -\frac{iR^+}{2}
\big)
\Big)_{AP}{\rm det'}^{1/2}\big(
-i\tau_2 
\partial_- 
\big)_{P}}{{\rm det'}^{1/2}\tau_2\big(2\partial_- \partial_+ -i{R^+}\partial_-\big)_{P}}
\eea
where $N_d$ is the standard normalization given by
\be
N_d= \frac{1}{(2\pi)^{d/2}}
\ee
and $AP$ and $P$ denotes respectively the anti-periodic and periodic boundary condition along the
Euclidean time circle direction. Let us first note that
\bea
\frac{{\rm det'}^{1/2}\big(
-i\tau_2 
\partial_- 
\big)_{P}}{{\rm det'}^{1/2}\big(
\partial_- 
\big)_{P}} =\left(\frac{i}{\tau_2}\right)^{d/2}
\eea 
This follows from the zeta function regularization
\be
\prod^\infty_{n=1}a= a^{\zeta(0)}=\frac{1}{\sqrt{a}}
\ee
where we used $\zeta(0)=-1/2$. Then 
\be
\prod_{n\neq 0} a =1/a
\ee

Let us compute the determinant
\be
{\rm det'}^{1/2} \big(2\tau_2 \partial_+ -i{\tau_2\, R^+}\big)_{P}
\ee
where 
\be
\partial_+ =\frac{1}{2} \big(
\partial_s +\partial_t
\big) =\frac{1}{2} \big(
\partial_s +i\partial_{t_E}
\big)
\ee
with $t= -i t_E$.
Note that
\be
s^- = s+i t_E = \tilde{s} + \tau \tilde{t} 
\ee
Then one finds
\be
\partial_+ =\frac{1}{2\tau_2} \big(
i \partial_ {\tilde{t}}-i \partial_{\tilde{s}}
\big) 
\ee
The eigenvalue of $\partial_+$ can be evaluated as
\be
\partial_+ \phi_{mn} =\frac{m+n \tau}{2\tau_2}\phi_{mn}
\ee 
with the eigenfunction
\be
\phi_{mn} = e^{-im {\tilde{t}}+i n {\tilde{s}}}
\ee
where $m,n \in \mathbb{Z}$ and we used the periodic boundary condition for the
time circle direction. For the antiperiodic boundary condition for the time circle direction, 
we get 
\be
\partial_+ \phi'_{mn} =\frac{m-1/2+n \tau}{2\tau_2}\phi'_{mn}
\ee
with
\be
\phi'_{mn} = e^{-i(m-1/2) {\tilde{t}}+i n {\tilde{s}}}
\ee
Since $R^{+ab}$ is an antisymmetric matrix transforming covariantly under the
$SO(4)$ rotation of the tangent space, this can be block diagonalized into the form
\bea
R^{+ab}=\left[
\begin{array}{cccc}
0& r_1& 0 & 0 \\
-r_1 & 0 & 0 &0 \\
0& 0& 0& r_2 \\
 0 & 0 &-r_2 &0
\end{array}\right]
\eea
by an appropriate $SO(4)$ rotation. We would like to evaluate then
\bea
D={\rm det}\left[\begin{array}{cc}
\partial_+  &  z \\
-z   & \partial_+
\end{array}\right]
\eea
Since one can use $ e^{+im {\tilde{t}}-i n {\tilde{s}}}$ as a basis instead of 
 $e^{-im {\tilde{t}}+i n {\tilde{s}}}$, we have 
\bea
D&=&{\rm det'}\left[\begin{array}{cc}
\partial_+  &  -z \\
z   & -\partial_+
\end{array}\right] = {\rm det'} \big(-\partial^2_+ +z^2) \big)\cr &=&
{\rm det'} \big(-\partial_+ -z) \, {\rm det'} \big(\partial_+ -z) = {\rm det'}^2 \big(\partial_+ -z)
\eea
Thus we need to evaluate 
\be
\prod' (m+ n\tau -z) =\prod \omega \prod' \Big(1-\frac{z}{\omega}\Big) 
\ee
where $\omega=m+ n\tau$. The computation of the former is fairly standard.  We use
\be
\prod (n+a) = -2 i \sin \pi a
\ee
leading to
\bea
\prod' \omega &=& \prod'(n) \prod_{m\neq 0}\prod_{m}(m+n \tau)=
-2\pi i \prod_{m\neq 0}(-2 i \sin \pi m \tau)
\cr
&=&2\pi \eta^2(\tau)
\label{zeta}
\eea
We now compute the second factor. Let us introduce the function $\sigma(\tau,z)$ defined by
\be
\sigma(\tau,z)=\prod' (1-z/\omega)e^{\Lambda(\omega)}
\ee 
where $\Lambda(z)$ is defined by
\be
\Lambda(z)=\frac{z}{\omega}+\frac{z^2}{2\omega^2}
\ee
We also introduce 
\be
\kappa(z)=(\ln \sigma(z))' = \frac{1}{z} +\sum'\left(
\frac{1}{z-\omega}+\frac{1}{\omega}+\frac{z}{\omega^2}
\right)
\ee
and the Weierstrass function
\be
{\cal P}(z)= -\kappa'(z) = \frac{1}{z^2} +\sum'\left(
\frac{1}{(z-\omega)^2}-\frac{1}{\omega^2}
\right)
\ee
We also introduce
\be
\sigma_1(z)=\frac{\sigma(z+1/2)}{\sigma(1/2)} e^{-\kappa_1 z}
= e^{-\frac{z^2}{2} e_1} \prod\Big(1-\frac{z}{\omega-1/2}\Big)e^{\Lambda_1(z)}
\ee
where $\kappa_1=\kappa(1/2)$, $e_1= {\cal P}(1/2)$ and
\be
\Lambda_1(z)=\frac{z}{\omega-1/2}+\frac{z^2}{2(\omega-1/2)^2}
\ee
It is well known that $\sigma(z)$ can be expressed in terms of the theta functions by 
\be
\sigma(z) =\frac{\theta_1(\tau,z)}{\theta'_1(\tau,0)} e^{\kappa_1 z^2}
\ee
Thus 
\be
\prod' \Big(1-\frac{z}{\omega}\Big)  = \frac{\sigma(z)}{z} e^{-\sum' \Lambda(z)}
=\frac{\theta_1(\tau,z)}{z \theta'_1(\tau,0)} e^{\kappa_1 z^2-\sum' \Lambda(z)}
\ee
where $e^{-\sum' \Lambda(z)}$ is the regularization term which can be potentially problematic, for instance, with symmetries
of the system.
Hence,
\be
\prod' \Big(\omega- z\Big)  
=\frac{\theta_1(\tau,z)}{z \eta(\tau)} e^{\kappa_1 z^2-\sum' \Lambda(z)}
\ee

Similarly for the fermion determinant with the anti-periodic boundary condition along the time circle direction, we need to evaluate
\be
\prod (m-1/2+ n\tau -z) =\prod (\omega-1/2) \prod  \Big(1-\frac{z}{\omega-1/2}\Big) 
\ee
With the zeta function regularization similar to (\ref{zeta}), one finds
\be
\prod (\omega-1/2) =\frac{\theta_2(\tau,0)}{\eta(\tau)}
\ee
Also, 
\be
\prod  \Big(1-\frac{z}{\omega-1/2}\Big) =\sigma_1(z) e^{\frac{z^2}{2}e_1 -\sum \Lambda_1(z)}
\ee
Noting 
\be
\sigma_1(z) =\frac{\theta_2(\tau,z)}{\theta_2(\tau,0)} e^{\kappa_1 z^2}
\ee
one finds
\be
\prod (\omega-1/2 -z)=\frac{\theta_2(\tau,z)}{ \eta(\tau)} e^{\kappa_1 z^2+\frac{z^2}{2}e_1-\sum \Lambda_1(z)}
\ee
where we use the cancellation of the regularization terms
\be
\kappa_1 z^2+\frac{z^2}{2}e_1-\sum \Lambda_1(z)=0
\ee
as a mathematical identity.

Therefore,
\be
\frac{\prod (\omega-1/2 -z)\Big)}{\prod' (\omega-z)}=\frac{z \theta_2(\tau,z)}{ \theta_1(\tau,z)} 
e^{\frac{z^2}{2}e_1-\sum \Lambda_1(z)+\sum' \Lambda(z)}=\frac{z \theta_2(\tau,z)}{ \theta_1(\tau,z)} 
\ee
and
\bea
Z(\tau,1/2)= \left(\frac{i}{2\pi \tau_2}\right)^2\int d^4 x \sqrt{g}\int \prod^4_{i=1} \psi^i_{0-} \prod^2_{k=1} 
\frac{z_k  \theta_2(\tau,z_k)}{ \theta_1(\tau,z_k)} 
\eea
where $z_k = i r_k \tau_2$. Integration of the fermion zero modes leads to
\bea
Z(\tau,1/2)=
\frac{1}{2} \left(
\left(\frac{\theta_3 (\tau,1/2)}{\theta_3 (\tau,0)}\right)^2 + \left(\frac{\theta_4 (\tau,1/2)}{\theta_4 (\tau,0)}\right)^2
\right)\tau(M_4)
\eea
where the signature is defined by
\be
\tau(M_4)=-\frac{1}{24 \pi^2} \int {\rm tr R^2}
\ee
for  the case of compact manifold.

\section{Derivation of Eq (\ref{Kawai}) using Hamiltonian quantization}\label{Kawaiapp}
Here we will quantize the oscillator modes using the Hamiltonian quantization but for the zero modes we use the path integral. For the path integral of the zero modes, we do not distinguish between space and time directions. We can have a constant fermionic zero mode in the path integral only if the fermion is periodic in both space and time directions. This is the case for the right-moving fermions only since the left-moving fermions are twisted by $y^{J_L}$. So only right-moving fermions $\psi^i_+$ have zero modes. 

Let us assume the world-sheet metric is Euclidean with complex coordinate $w$. Then the sigma model action becomes
\bea
S &=& \int dt_E\int \frac{ds}{2\pi} \(2 \partial_w X^i \partial_{\bar{w}} X^i + i \psi^i_+ D_w \psi^i_+ - i \psi^i_- D_{\bar{w}} \psi^i_- + \frac{1}{4} R_{ijkl} \psi^i_+ \psi^j_+ \psi^k_- \psi^l_-\)
\eea
where the world-sheet is the torus 
\bea
w &\sim & w + 2\pi\tau\cr
w &\sim & w + 2\pi
\eea
and we have the eigenvalues
\bea
\partial_w &=& \frac{1}{2\tau_2}\(n-\bar{\tau}m\) 
\eea
where $m$ and the RR integer mode numbers, and $n$ are integer mode numbers for the time direction to be used in the path integral. For the reduction to quantum mechanics we pick the RR mode number $m=0$ sector (rigid string).

Expaning the action in Riemann normal coordinates and defining 
\bea
R_{ij} &=& \frac{1}{2} R_{ijkl} \psi^{k}_{0+} \psi^{l}_{0+}
\eea
we get
\bea
S &=& \int dt_E \int \frac{ds}{2\pi} \(-2 X^i \(\partial_{\bar{w}} \delta_{ij} + \frac{1}{2} R_{ij} \) \partial_w X^j + \psi^i_- \(\partial_{\bar{w}} \delta_{ij} + \frac{1}{2} R_{ij} \) \psi^j_-
- \psi^i_+ \partial_w \psi^i_+\)
\eea
For the oscillator modes we will now obtain the Hamiltonian. We begin with the fermions for which we find the left and right moving Hamiltonians
\bea
H_L &=& \sum_{n=1}^{\infty} \(n \((\psi^i_{n,-})^{\dag} \psi^i_{n,-} - \frac{1}{2}\) - \frac{1}{2} R_{ij} (\psi^i_{n,-})^{\dag} \psi^j_{n,-}\)\cr
H_R &=& \sum_{n=1}^{\infty} \(n \(\psi^i_{n,+} (\psi^i_{n,+})^{\dag} - \frac{1}{2}\) - \frac{1}{2} R_{ij} (\psi^i_{n,-})^{\dag} \psi^j_{n,-}\)
\eea
where the canonical commutation relations imply that for any matrix $M_{ij}$ and operator
\bea
H_n &=& M_{ij} (\psi^i_{n,-})^{\dag} \psi^j_{n,-}
\eea
we have the following commutation relation
\bea
[H_n,(\psi^i_{n,-})^{\dag}] &=& M_{ij} (\psi^i_{n,-})^{\dag}
\eea
To find the eigen-energies of the Hamiltonians we thus just need to find the eigenvalues of the corresponding matrices. We off-diagonalize $R_{ij}$ with off diagonal eigenvalues $x_a$. Let us assume target space is a four-manifold. Then $a=1,2$ and
\bea
R_{ij} &=& \(\begin{array}{cccc}
0 & x_1 & 0 & 0\\
-x_1 & 0 & 0 & 0\\
0 & 0 & 0 & x_2\\
0 & 0 & -x_2 & 0
\end{array}\)
\eea
We also have
\bea
J_L &=& \(\begin{array}{cccc}
0 & i & 0 & 0\\
-i & 0 & 0 & 0\\
0 & 0 & 0 & i\\
0 & 0 & -i & 0
\end{array}\)
\eea
The energy eigenvalues now obtained as 
\bea
H_L &=& \(n \pm \frac{ix_a}{2}\) N_n^L\cr
H_R &=& nN_n^R \pm \frac{ix_a}{2} N^L_n
\eea
where $N_n^{L,R} = 0,1$ are the fermionic number operators in left and right sectors. (We ignore the zero point energy contribution, which in the end will cancel against the bosons' zero point energies.) We then get the fermionic oscillators contribution 
\bea
Z_F &=& \prod_{a}  \prod_{n=1}^{\infty} (1 - q^n e^{ix_a} y) (1 - q^n e^{-ix_a} y^{-1})  (1 - \bar{q}^n)^2 
\eea
where we rescaled $x_a$ to $-2\pi \tau_2 x_a$. By supersymmetric pairing of all the oscillator modes it is immediately clear that the oscillator mode contribution to the elliptic genus, by including the bosons, is given by
\bea
Z_{osc}(q,y) &=& \prod_a \prod_{n=1}^{\infty} \frac{(1-q^n e^{ix_a} y) (1 - q^n e^{-ix_a} y^{-1})}{(1-q^n e^{ix_a}) (1-q^n e^{-ix_a})}
\eea
For the zero modes we consider the rigid string sector for which we have the quantum mechanics action
\bea
S_0 &=& \frac{1}{2}\int dt \( X^i \(i\partial_t \delta_{ij} + \tau_2 R_{ij} \) \frac{i}{\tau_2} \partial_t X^j + \psi^i_- \(i\partial_t \delta_{ij} + \tau_2 R_{ij} \) \psi^j_-
+ \psi^i_+ i\partial_t \psi^i_+\)
\eea
The  contribution from the zero modes can be computed from the path integral 
\bea 
Z_0(y) &=& \int D\psi DX e^{-S_0}
\eea
with twisted boundary conditions 
\bea
\psi^i_-(t+2\pi) &=& \(e^{2\pi i z J_L}\)^{ij} \psi^j_-(t)\cr
\psi^i_+(t+2\pi) &=& \psi^i_+(t)
\eea
We then get
\bea
Z_0(y) &=& \(\frac{\prod_{n\in \mb{Z}} \det\(-n\delta_{ij} + z J_{L,ij} + \tau_2 R_{ij}\)}{\prod_{n\neq 0}\det\(-n\delta_{ij} + \tau_2 R_{ij}\)}\)^{\frac{1}{2}}
\eea
We now use the zeta function regularization to get
\bea
Z_0(y) &=& \prod_{a=1,2} \frac{x_a \sin (\pi z-ix_a)}{\sinh x_a}
\eea
The full elliptic genus, by including both oscillators and zero modes can be expressed as
\bea
Z(q,y) &=& \int_{M_{2n}} \prod_{a=1,2} x_a \frac{\theta_1(q,y e^{2x_a})}{\theta_1(q,e^{2x_a})}
\eea
As a special case, we find the Euler characteristic at $y=1$,
\bea
Z(q,1) = \int_M \prod_a x_a = \chi(M)
\eea
We see that the limit $y\rightarrow 1$ is smooth. A direct computation at $y=1$ requires additional fermionic zero modes to be taken into account because there is twisting of fermionic boundary conditions at this point. 

Also in the limit $q\rightarrow 0$ and at the point $y=-1$ we find
\bea
Z(0,-1) = \int_M \prod_a \frac{x_a}{\tanh x_a} = \sigma(M)
\eea
which is the Hirzebruch signature. 

We may also notice that in the flat space limit $x_a \rightarrow 0$ the integrand formally corresponds to the flat space elliptic genus Eq (\ref{u1}) with chemical potential $u=1$, although this limit is singular. In fact, the elliptic genus at the point $u=1$ on $\mb{R}^4$ is zero due to extra left-moving fermionic zero modes.

\section{Application of formula (\ref{general})}\label{generalapp}
Let us first apply the formula (\ref{general}) to $k=1$ for generic $N$. Let us assume that we have a Young diagram at $i=1$ with one box, and no Young diagrams at $i=2,\cdots,N$. Then
\bea
E_{ij}(1,1) &=& \left\{\begin{array}{ll}
0 & {\mbox{if $j=i$}}\\
\mu_{ij} - i(\gamma_1+\gamma_R) & {\mbox{if $j\neq i$}}
\end{array}\right.
\eea
We then get
\bea
I_{\{Y_1,\cdots,Y_N\}, Y_i\neq \emptyset} &=& I_{com}(\gamma_1,\gamma_2,\gamma_R) \prod_{j\neq i}^N I(\mu_{ij},\gamma_2,\gamma_R)
\eea
where we introduce the quantities
\bea
I_{com}(\gamma_1,\gamma_2,\gamma_R) &=& \frac{\sin \frac{\gamma_1+\gamma_2}{2}\sin \frac{\gamma_1-\gamma_2}{2}}{\sin \frac{\gamma_1+\gamma_R}{2}\sin \frac{\gamma_1-\gamma_R}{2}}\cr
I(\mu_{ij},\gamma_2,\gamma_R) &=& I_{com}(\gamma_R+i\mu_{ij},\gamma_2,\gamma_R)
\eea
following the notation of \cite{Kim:2011mv}. We see that $I_{com}$ diverges as $\gamma_1,\gamma_R\rightarrow 0$. 

\subsection{Series expansions for $N=3$}
The relation between the multi-particle and single-particle indices is quite simple if we just expand up to first order in $q$, the complication starts at quadratic order. After we have extracted the divergent $I_{com}$, we can put $\gamma_1=\gamma_R=0$. In terms of the  basic building block
\bea
I(x_{ij},y) &=& \frac{(1-y^{-1}x_{ij})(1-yx_{ij})}{(1-x_{ij})^2}\cr
&=& 1 + \frac{(1-y)^2}{y}\sum_{n=1}^{\infty} n x_{ij}^n
\eea
the single particle indices are given by
\bea
z'_{N=2}(x_{12},y) &=& \(I(x_{12})+I(x_{21})\)q + \Ordo(q^2),\cr
z'_{N=3}(x_{ij},y) &=& \(I(x_{12}) I(x_{13}) + I(x_{21})I(x_{23}) + I(x_{31})I(x_{32})\)q + \Ordo(q^2)
\eea
From the $\Ordo(x_{12}^0)$ term we read off the index coming from having no M2 branes stretching between the two M5's. Yet we can have a wave with momentum $k$ in each M5 brane. This amounts to the index of two copies of $U(1)$ MSYM theory. For each copy, this index was found to be $1$ for each $k=1,2,3,\cdots$. Thus we shall have
\bea
z'_{N=2} &=& 2(q+q^2+q^3+\cdots) + \Ordo(x_{12})
\eea
At the first few orders we find
\bea
z'_{N=2,n=0} &=& 2q+ \cdots\cr
z'_{N=2,n=1} &=& 1+2(2-y-y^{-1})q+\cdots\cr
z'_{N=2,n\geq 2} &=& 0+2n(2-y-y^{-1})q+\cdots
\eea
For $n=1$ we have added $1$ by hand. This corresponds to a single M2 brane (or W-boson) stretched between the two M5's which carries zero longitudinal momentum (or zero instanton number).  For $n=1$ there is a closed form of the index given by \cite{Kim:2011mv}
\bea
z_{N=2,n=1} &=& -\frac{\theta_1(q,yu_1)\theta_1(q,yu_1^{-1})}{\theta_1(q,u_1)^2}
\eea
Here we keep the chemical potential $u_1 = e^{i\gamma_1}$ before we have extracted the divergent $I_{com}(y,u_1)$ out of it. If we extract this divergent piece, then we can take $\gamma_1 \rightarrow 0$ in the remaining piece, which then will have the series expansion 
\bea
z'_{N=2,n=1} &=& 1 - \frac{2(1-y)^2}{y}q\cr
&&+\frac{(1-y)^2(1-(8-y)y)}{y^2}q^2 + \frac{8(1-y)^2(1-(3-y)y)}{y^2}q^3 + \Ordo(q^4)
\eea
At $y=-1$ we have
\bea
z'_{N=2,n=1}(q,-1) &=& 1 + 8q + 40 q^2 + 160 q^3 + \Ordo(q^4)
\eea
For $N=3$ we may extract the single particle index associated to an M2 brane stretching between M5$_1$ and M5$_3$ by extracting the coefficient of $x_{13}$ keeping in mind that $x_{12} x_{23} = x_{13}$. This coefficient is found to be 
\bea
z'_{N=3,n=1}(q,y) &=& 1+ \(10-6(y+y^{-1})+y^2+y^{-2}\) q + \Ordo(q^2)
\eea
This we can also express in the form
\bea
z'_{N=3,n=1}(q,y) &=& z'_{N=2,n=1}(q,y) Z_0(q,y) 
\eea
where
\bea
Z_0(q,y) &=& 1+ \frac{(1-y)^4}{y^2} q + \Ordo(q^2)
\eea
It also turns out that 
\bea
z'_{N=N,n=1}(q,y) &=& z'_{N=2,n=1}(q,y)\(Z_0(q,y)\)^{N-2}
\eea
for $N=3,4,5,\cdots$, which thus all can be expressed in terms of the universal factor $Z_0(q,y)$. For $y=-1$ this universal factor has the expansion
\bea
Z_0(q,-1) &=& 1+16 q + 96 q^2 + 448 q^3 +\cdots
\eea
Let us also note that 
\bea
z'_{N=2,n=2}(q,-1) &=& 0 + 16q + 288 q^2 + 2880 q^3+\cdots
\eea
In the next subsection we will spend some effort on obtaining the general $y$-dependence of $z'_{N=2,n=2}(q,y)$ up to quadratic order. 

\subsection{Expansion up to quadratic order in $q$ for $N=2$}
Going to the next higher order in $q$ enables us to make a more convincing check of our result and of TST duality. In this subsection we will therefore expand the index to quadratic order in $q$ but for simplicity we will just consider the case $N=2$. We will extract a single particle index to the same order in $q$. Explicit expressions for the indices at these first few orders in $q$ can be inferred from the general formula (\ref{general}) as
\bea
I_0 &=& 1\cr
\frac{I_1}{I_{com}(\gamma)} &=& I(\mu,\gamma)+I(-\mu,\gamma)\cr
\frac{I_2}{I_{com}(\gamma)} &=& 4I_{com}(\gamma_R) I(\mu,\gamma) I(\mu-i\gamma_R,\gamma)\cr
&&+I_{com}(\gamma)I(\mu+i\gamma_R,\gamma)I(-\mu+i\gamma_R,\gamma)
\eea
Here we use the short-hand notation $\gamma$ for $(\gamma_2,\gamma_R)$ and we write $\mu$ in place of $\mu_{12}$ and we define $x = e^{-\mu}<1$.

Let us now expand to quadratic order 
\bea
z'(q,\mu,\gamma) &=& q z_1(\mu,\gamma) + q^2 z_2(\mu,\gamma) + \Ordo(q^3)
\eea
and
\bea
I &=& 1+qI_{com}(\gamma) z_1(\mu,\gamma)\cr
&&+q^2 \(I_{com}(\gamma) z_2(\mu,\gamma) + \frac{1}{2} \(I_{com}(2\gamma)z_1(2\mu,2\gamma)+I_{com}(\gamma)^2 z_1(\mu,\gamma)^2\) \) + \Ordo(q^3)
\eea
Matching this result with the expression we know for the multi-particle index, we get
\bea
z_1(\mu,\gamma) &=& \frac{I_1}{I_{com}(\gamma)}\cr
z_2(\mu,\gamma) &=& \frac{I_2}{I_{com}(\gamma)} - \frac{1}{2} \(\frac{I_{com}(2\gamma)}{I_{com}(\gamma)} z_1(2\mu,2\gamma)+I_{com}(\gamma)z_1(\mu,\gamma)\)
\eea
Explicitly we find 
\bea
z_1(x,y) &\rightarrow & \frac{2(1-yx)(1-y^{-1}x)}{(1-x)^2}
\eea
in the limit $\gamma_R \rightarrow 0$. The expression has the series expansion
\bea
z_1(x,y) &=& 2 + \(4-2(y+y^{-1}\)x + \(8-4(y+y^{-1})\)x^2+\Ordo(x^3)
\eea
We could of course substitute $z_1$ for $I_1$ in the expression for $z_2$. However we like to keep this form as we will take the limit $\gamma_R \rightarrow 0$ in which $I_{com}(\gamma)$ diverges while $z_1(\mu,\gamma)$ is finite. Furthermore we have the finite limit (under which $y=e^{i\gamma_2}$ is kept fixed)
\bea
\frac{I_{com}(2\gamma)}{I_{com}(\gamma)} \rightarrow \frac{1}{4}\cdot\frac{2-y^2-y^{-2}}{2-y-y^{-1}}
\eea
We then note that both $\frac{I_2}{I_{com}(\gamma)}$ and $-\frac{1}{2}I_{com}(\gamma)z_1(\mu,\gamma)$ diverges. Happily the divergences precisely cancel so that the sum has a finite limit
\bea
\frac{I_2}{I_{com}(\gamma)} - \frac{1}{2}I_{com}(\gamma)z_1(\mu,\gamma) \rightarrow i_2(x,y)
\eea
We have obtained a closed expression for $i_2(x,y)$ using Mathematica, but it is quite long. The expression simplifies at $y=-1$ to
\bea
i_2(x,-1) &=& \frac{2(1+x)^2(1+12x+14x^2+12x^3+x^4)}{(1-x)^6} 
\eea
For generic $y$ it has the series expansion
\bea
i_2(x,y) &=& \frac{5}{2} + \frac{1}{4}(y+y^{-1}) + \(18-10(y+y^{-1})+y^2+y^{-2}\)x \cr
&&+ \(113 - \frac{287}{4} (y+y^{-1}) + \frac{31}{2} (y^2+y^{-2}) - \frac{1}{4} (y^3+y^{-3})\)x^2 + \Ordo(x^3)
\eea
We now obtain $z_2$ as
\bea
z_2(x,y) &=& i_2(x,y) - \frac{1}{8} \cdot \frac{2-y^2-y^{-2}}{2-y-y^{-1}} z_1(x^2,y^2)
\eea
where all quantities on the right hand side are now finite. We may notice that the second term vanishes at $y=-1$. For generic $y$ we get the following series expansion,
\bea
\frac{1}{8} \cdot \frac{2-y^2-y^{-2}}{2-y-y^{-1}} z_1(x^2,y^2) &=& \frac{(1+y)^2}{4y} - \frac{(1-y)^2(1+y)^4}{4y^3}x^2 + \Ordo(x^4)
\eea
and then we get
\bea
z_2(x,y) &=& 2 + \(18-10(y+y^{-1})+y^2+y^{-2}\)x \cr
&&+  \(112-72(y+y^{-1})+16(y^2+y^{-2})\) x^2 + \Ordo(x^3)
\eea

From these results we can read off the contribution coming from a single and from two M2 branes stretching between two M5 branes by reading off the coefficient of $x$ and $x^2$ respectively, 
\bea
z'_{N=2,n=1}(q,y) &=& 1 + \(4-2(y+y^{-1}\)q + \(18-10(y+y^{-1})+y^2+y^{-2}\) q^2 + \Ordo(q^3)\cr
z'_{N=2,n=2}(q,y) &=& \(8-4(y+y^{-1})\) q + \(112 -72(y+y^{-1})+16(y^2+y^{-2})\) q^2 \cr
&& + \Ordo(q^3)
\eea
It should be noted that the series expansion for $z'_{N=2,n=1}$ agrees with the expansion of the proposed closed formula (\ref{su2}). 

These expansions when evaluated at $y=-1$ read \cite{Kim:2011mv}
\bea
z'_{N=2,n=1}(q,-1) &=& 1 + 8 q + 40 q^2 + \Ordo(q^3)\cr
z'_{N=2,n=2}(q,-1) &=& 0 + 16q + 288 q^2 + \Ordo(q^3)
\eea

\section{Spectral flow for K3 elliptic genus}
Here we confirm that the spectral flow method reproduces the known elliptic genus for the K3 target space. Here we have $\chi(K3) = 24$ and $\sigma(K3) = 16$. We then use the following boundary data 
\bea
&& Z(\tau,0)= 24\cr
&&
Z\Big(\tau, \frac{1}{2}\Big)= 8 \left(
\left(\frac{\theta_3 (\tau,1/2)}{\theta_3 (\tau,0)}\right)^2 + \left(\frac{\theta_4 (\tau,1/2)}{\theta_4 (\tau,0)}\right)^2
\right) 
\eea
and spectral flow, to find a series expansion for generic $z$, 
\bea
Z(\tau,z)&=& 2\left[ \big(10+(y+y^{-1})\big)+ \big(108+10(y^2+ y^{-2})-64(y+y^{-1})\big)q \right.
\cr
&+&\left. \big(808+(y^3+y^{-3})+108(y^2+ y^{-2})-513(y+y^{-1})\big)q^2 + \cdots \right]
\eea
This now agrees with the known expression for the K3 elliptic genus \cite{Kawai:1993jk},
\bea
Z(\tau,z)&=&8 \left( \left(\frac{\theta_2 (\tau,z)}{\theta_2 (\tau,0)}\right)^2+
\left(\frac{\theta_3 (\tau,z)}{\theta_3 (\tau,0)}\right)^2 + \left(\frac{\theta_4 (\tau,z)}{\theta_4 (\tau,0)}\right)^2
\right) 
\eea
when this is series expanded. This is a weak Jacobi form of weight zero index one which is fully covariant under the SL(2,$\mathbb{Z}$) modular transformation.

We can also apply the same technique on the refined elliptic genus for K3 target space where this leads to
\bea
\h Z^{K3} (\tau,z,\mu)&=& 8 \left(
\frac{\theta_2 (\tau,z+\mu)\theta_2 (\tau,z-\mu)}{\theta^2_2 (\tau,\mu)}+
\frac{\theta_3 (\tau,z+\mu)\theta_3 (\tau,z-\mu)}{\theta^2_3 (\tau,\mu)} \right. \cr
&+&  \left.
\frac{\theta_4 (\tau,z+\mu)\theta_4 (\tau,z-\mu)}{\theta^2_4 (\tau,\mu)}
\right) 
\label{regk3}
\eea
which has an expansion
\be
\h Z^{K3} (\tau,z,\mu)= 12+4\frac{\big(y+y^{-1}-2\big)x}{(1-x)^2}   + \Ordo(q)
\ee
The above refined elliptic genus is fully covariant under the SL(2,$\mathbb{Z}$). By choosing an appropriate U(1) which is invariant under the spectral flow, one can verify the above computation directly by the orbifold computation of $T^4/\mathbb{Z}_2$.

\end{document}